\documentclass[12pt,preprint]{aastex}

\usepackage{emulateapj5}
\usepackage{apjfonts}
\usepackage{epsf}

\newcommand{\bm}[1]{\mbox{\boldmath$#1$}}

\slugcomment{Submitted to ApJ; Draft version \today}

\shorttitle{Cosmic shear statistics in the Suprime-Cam 2.1 sq deg field}
\shortauthors{T. Hamana et al.}

\begin{document}
\title{Cosmic shear statistics in the Suprime-Cam 2.1 sq deg field:\\
Constraints on $\Omega_{\rm m}$ and $\sigma_8$\altaffilmark{1}}

\author{
Takashi Hamana\altaffilmark{2,3},
Satoshi Miyazaki\altaffilmark{2}, 
Kazuhiro Shimasaku\altaffilmark{4}, 
Hisanori Furusawa\altaffilmark{2},
Mamoru Doi\altaffilmark{5}, 
Masaru Hamabe\altaffilmark{6}, 
Katsumi Imi\altaffilmark{6},
Masahiko Kimura\altaffilmark{8}, 
Yutaka Komiyama\altaffilmark{2}, 
Fumiaki Nakata\altaffilmark{2},
Norio Okada\altaffilmark{2}, 
Sadanori Okamura\altaffilmark{4},
Masami Ouchi\altaffilmark{4}, 
Maki Sekiguchi\altaffilmark{8},
Masafumi Yagi\altaffilmark{2} and
Naoki Yasuda\altaffilmark{2}
}

\email{hamana@iap.fr}

\altaffiltext{1}{Based on data collected at Subaru Telescope, 
which is operated by the National Astronomical Observatory of Japan.}
\altaffiltext{2}{National Astronomical Observatory of Japan, 
Mitaka, Tokyo 181-8588, Japan}
\altaffiltext{3}{Institut d'Astrophysique de Paris, 98bis Boulevard Arago, 
F 75014 Paris, France}
\altaffiltext{4}{Department of Astronomy, University of Tokyo, 
Bunkyo, Tokyo 113-0033, Japan}
\altaffiltext{5}{Institute of Astronomy, University of Tokyo, 
Mitaka, Tokyo 181-0015, Japan}
\altaffiltext{6}{Department of Mathematical and Physical Sciences,
Japan Women's University, Bunkyo, Tokyo 112-8681, Japan}
\altaffiltext{7}{Communication Network Center (Tsu-den) , 
Mitsubishi Electric, Amagasaki, Hyogo 661-8661, Japan}
\altaffiltext{8}{Institute for Cosmic Ray Research, 
University of Tokyo, Kashiwa,Chiba 277-8582, Japan}

\begin{abstract}
We present measurements of the cosmic shear correlation in the shapes of 
galaxies in the Suprime-Cam 2.1 deg$^2$ $R_c$-band imaging data.
As an estimator of the shear correlation originated from the gravitational 
lensing, we adopt the aperture mass variance, which most naturally decomposes
the correlation signal into E and B (non-gravitational lensing) modes.
We detect a non-zero E mode variance on scales between
$\theta_{ap}=2\arcmin$ and 40\arcmin.
We also detect a small but non-zero B mode variance on scales
larger than $\theta_{ap}>5\arcmin$.
We compare the measured E mode variance to the model predictions in CDM
cosmologies using maximum likelihood analysis.
A four-dimensional space is explored, which examines $\sigma_8$,
$\Omega_{\rm m}$, $\Gamma$ (the shape parameter of the CDM power
spectrum) and $\bar{z}_s$ (a mean redshift of galaxies).
We include three possible sources of error: statistical noise,
the cosmic variance estimated using numerical experiments,
and a residual systematic effect estimated from the B mode variance.
We derive joint constraints on two parameters by marginalizing over
the two remaining parameters.
We obtain an upper limit of $\Gamma<0.5$ for $\bar{z}_s>0.9$
(68\% confidence).
For a prior $\Gamma\in[0.1,0.4]$ and $\bar{z}_s\in[0.6,1.4]$,
we find
$\sigma_8=(0.50_{-0.16}^{+0.35})\Omega_{\rm m}^{-0.37}$ for
$\Omega_{\rm m}+\Omega_{\Lambda}=1$ and
$\sigma_8=(0.51_{-0.16}^{+0.29})\Omega_{\rm m}^{-0.34}$ for
$\Omega_{\Lambda}=0$ (95\% confidence).
If we take the currently popular $\Lambda$CDM model ($\Omega_{\rm
m}=0.3$,
$\Omega_{\lambda}=0.7$, $\Gamma=0.21$), we obtain a one-dimensional
confidence interval on $\sigma_8$ for the 95.4\% level, 
$0.62<\sigma_8<1.32$ for $\bar{z}_s\in[0.6,1.4]$.
Information on the redshift distribution
of galaxies is key to obtaining a correct cosmological constraint.
An independent constraint on $\Gamma$ from other
observations is useful to tighten the constraint.
\end{abstract}


\section{Introduction}
Cosmic shear, that is coherent distortions in distant galaxy images
resulting
from weak gravitational lensing by large-scale structures, has been
recognized as a powerful tool for cosmology because it directly
probes
matter distribution regardless of any relation between mass and
light (for reviews see, Mellier 1999, Bartelmann \& Schneider 2001).
Since the first reports of the detection of cosmic shear correlations
(Van Waerbeke et al., 2000; Witteman et al., 2000; Bacon, Refregier \&
Ellis 2000; Kaiser, Wilson \& Luppino 2000; Maoli et al.~2001; ), cosmic
shear statistics have become
a promising probe of cosmological parameters.
Indeed, recent measurements of cosmic shear correlation have put useful
constraints on the matter density parameter $\Omega_{\rm m}$ and
the matter power spectrum normalization $\sigma_8$
(Maoli et al.~2001; Bacon et al.~2002; Van Waerbeke et al.~2001; 2002;
Hoekstra et al.~2002a; Hoekstra, Yee \& Gladders 2002b; Brown et
al.~2003; Jarvis et al.~2003).
As the cosmic shear correlation is primarily sensitive to density
fluctuation at intermediate redshifts ($0.2<z<0.7$) and on scales from 
quasi-linear to
nonlinear ($1<\theta<50$ arcmin corresponding to comoving scales of
$0.38<r<19.2h^{-1}$Mpc at $z=0.5$ for a cosmological model with
$\Omega_{\rm m}=0.3$ and $\Omega_\Lambda=0.7$),
it provides cosmological information that is independent of
other observations, such as galaxy clustering, cluster number counts,
and the cosmic microwave background anisotropies, and thus is
complementary to these techniques.

Cosmic shear measurement is not an easy task; it requires
a rigorous observation strategy, specifically, a deep and wide-field
survey with very high image quality.
A large number density of distant galaxies, ideally
$n_g>30$arcmin$^{-2}$,
is needed to suppress random noise due to the intrinsic galaxy
ellipticity. A wide survey area is necessary for
suppression of the cosmic variance.
In addition, good seeing conditions are essential for precise
measurements of a galaxy's shape.
Suprime-Cam, a wide-field camera mounted on the prime focus of the
8.2-m Subaru telescope, is an almost ideal camera for a weak lensing
survey.
It has a field of view of $34\arcmin\times27\arcmin$ with
$0\arcsec.202$ pixel$^{-1}$.
The median seeing in the $I_c$ band, monitored over a period of 
one-and-a-half years,
is reported to be $\sim$ 0.6 arcsec (Miyazaki et al. 2002b).
Subaru's light-gathering power enables a complete magnitude of
$R_C = 25.2$ (the turn-around of galaxy counts) to be obtained by a
30-min exposure, which provides a galaxy number density of
$n_g \sim 33$arcmin$^{-2}$, after object selection (see \S 3).
These advantages allow a weak lensing survey to be carried out
efficiently.

In this paper we present the results of our analysis of
$R_c$-band imaging data from a Suprime-Cam 2.1 deg$^2$ field.
This field is composed of nine pointings in a $3\times 3$ mosaic
configuration which covers a contiguous $1.64\arcdeg\times1.28\arcdeg$
area.
A great advantage of these data is their very good and homogeneous
image quality over the whole field, which allows us to obtain secure
weak lensing measurements.
These data were also used for the first halo number counts using the
weak lensing technique (Miyazaki et al., 2002a), which demonstrated that
a weak lensing halo survey is indeed a promising way to study the
distribution and evolution of large-scale structures. 
In this paper, we concentrate on the measurement of cosmic shear
correlations caused by large-scale structures.
We carry out a full maximum likelihood analysis of the cosmic shear
correlation over four parameters, $\Omega_{\rm m}$, $\sigma_8$,
$\Gamma$ (a shape parameter of the Cold Dark Matter power spectrum) and
$\bar{z}_s$ (the mean redshift of the source galaxy distribution).
We derive joint constraints on two parameters by marginalizing over
the two remaining parameters, and obtain a constraint on $\Omega_{\rm m}$
and $\sigma_8$. We also obtain an upper limit on $\Gamma$.
Confidence intervals on $\sigma_8$ for the currently popular $\Lambda$CDM
model ($\Omega_{\rm m}=0.3$, $\Omega_{\Lambda}=0.7$ and $\Gamma=0.21$)
are also derived.

The outline of this paper is as follows.
In \S 2 we briefly discuss the theory of cosmic shear statistics and
summarize the analytical formulas that are used to compute the
theoretical predictions. Details of the observations and data are
described in \S 3. A galaxy shape analysis is presented in \S 4.
Measurement of the cosmic shear correlation is presented in \S 5.
The measured cosmic shear correlation signal is compared with the
theoretical prediction using maximum likelihood analysis in \S 6.
Finally, \S 7 presents a summary and discussion.
In the Appendix, we discuss tests of the anisotropic point spread
function correction procedure in some detail.

\section{Basics of cosmic shear correlation}

In this section, we provide a basic description of the theory of the cosmic
shear correlation (see the reviews by Mellier 1999; Bartelmann \&
Schneider 2001 for details).

The observable shear two-point statistics can be related to the
convergence power spectrum defined by
\begin{equation}
\label{eq:Pkappa}
P_\kappa(l) = {{9 \Omega_{\rm m}^2}\over 4}\left({{H_0}\over c}\right)^4
\int_0^{\chi_H}d\chi~\left[ {{W(\chi)}\over a(\chi)}\right]^2
P_\delta \left[ {l \over {f_K(\chi)}};\chi \right],
\end{equation}
where $H_0$ is the Hubble parameter, $\chi$ is the radial comoving
distance,
$\chi_H$ corresponds to the
horizon, $a(\chi)$ is the scale factor and $f_K(\chi)$ is the comoving
angular diameter distance. $P_\delta(k)$ is the matter power
spectrum, for which we adopt the fitting function of the CDM power
spectrum given by Bardeen et al.~(1986), but we treat the shape parameter,
$\Gamma$, as a free parameter.
To take into account the effect of nonlinear growth of the
density field, which has a significant impact on the shear correlation
function on scales below one degree (Jain \& Seljak 1997), we use the
fitting function of Peacock \& Dodds (1996).
$W(\chi)$ is the source weighted distance ratio given by,
\begin{equation}
\label{eq:W}
W(\chi)= \int_\chi^{\chi_H} d\chi'~n_s(\chi')
{{f_K(\chi'-\chi)}\over {f_K(\chi)}},
\end{equation}
here $n_s(\chi)$ is the normalized redshift distribution of source
galaxies, which we discuss later (\S 6.1).

Several real space estimators of the shear correlation have been proposed,
including: the shear two-point correlation function
(Blandford et al. 1991; Miralda-Escude 1991; Kaiser 1992),
the top-hat shear variance (Bernardeau, Van Waerbeke \& Mellier 1997)
and the aperture mass variance (Schneider et al. 1998).
For our maximum likelihood analysis, we adopt the aperture mass variance,
which has the very useful property that it naturally carries out
E/B mode decomposition (Schneider et al. 1998; Crittenden et al. 2002;
Pen, Van Waerbeke \& Mellier 2002).
As gravitational lensing produces only E mode shear, E/B mode
decomposition allows contamination from B mode shear (which
is not caused by gravitational lensing) to be suppressed,
and the amplitude of the B mode variance can be used to estimate the
amplitude of the systematic error.
The aperture mass is defined by
\begin{equation}
\label{eq:Map}
M_{ap}=\int d^2\theta~U(\theta)\kappa(\bm{\theta}),
\end{equation}
where $\kappa(\theta)$ is the lensing convergence field,
and $U(\theta)$ is the compensated filter, for which we adopt the
following function proposed by Schneider et al. (1998)
\begin{equation}
\label{eq:U}
U(\theta)={9 \over {\pi \theta_{ap}^2}}
\left( 1- {{\theta^2}\over {\theta_{ap}^2}}\right)
\left( {1\over3}- {{\theta^2}\over {\theta_{ap}^2}}\right),
\end{equation}
for $\theta<\theta_{ap}$, and 0 otherwise.
It should be emphasized that this filter probes an effective scale of
$\theta_{ap}/5$ not $\theta_{ap}$.
The aperture mass can be calculated directly from the tangential shear
$\gamma_t$ (the tangential component of shear defined in the local frame
connecting the aperture center to a galaxy), without the need for a mass
reconstruction, by
\begin{equation}
\label{eq:Map2}
M_{ap}=\int d^2\theta~Q(\theta)\gamma_t(\theta),
\end{equation}
where $Q(\theta)$ is given from $U(\theta)$:
\begin{equation}
\label{eq:Q}
Q(\theta)={2\over {\theta^2}} \int d\theta'~U(\theta')-U(\theta).
\end{equation}
The aperture mass variance is related to the convergence power spectrum
eq.~(\ref{eq:Pkappa}) by
\begin{equation}
\label{eq:Mapvar}
\langle M_{ap}^2 \rangle(\theta_{ap}) =
2\pi \int dl~l~P_\kappa(l)
\left[ {{12} \over {\pi (l\theta_{ap})^2}} J_4(l\theta_{ap})\right]^2,
\end{equation}
where $J_4$ is a fourth-order Bessel function of the first kind.

We compute the aperture mass variances from the shear
correlation functions using relations eqs.~(\ref{MapVarE}) and
(\ref{MapVarB}).
This approach has some advantages over direct measurement in that
(i) it is the least affected by defects in the data, such as masking by bright
stars, (ii) and it does not depend on the geometry of the data;
(iii) thus it uses all information in the data (Hoekstra et al. 2002b).
The two shear two-point correlation functions that are measured are:
\begin{equation}
\label{eq:xitt}
\xi_{tt}(\theta) =
{{\sum_{i,j}^{N_s} w_i w_j \gamma_{t,i} \gamma_{t,j} }
\over
{\sum_{i,j}^{N_s} w_i w_j}},
\end{equation}
and
\begin{equation}
\label{eq:xirr}
\xi_{rr}(\theta) =
{{\sum_{i,j}^{N_s} w_i w_j \gamma_{r,i} \gamma_{r,j} }
\over
{\sum_{i,j}^{N_s} w_i w_j}},
\end{equation}
where $\theta=|\bm{x_i}-\bm{x_j}|$, $N_s$ is the number of source galaxies,
$\gamma_{t}$ and $\gamma_{r}$ are the tangential
and 45$\arcdeg$ rotated shear in the frame connecting the pair of
galaxies,
and $w$ is a weight that expresses the reliability of the shape
measurement for each galaxy (discussed in \S 4).
For the following discussion, it is useful to define, $\xi_+(\theta)$
and $\xi_-(\theta)$, which are the sum and difference of the shear
two-point correlation functions defined by eqs.~(\ref{eq:xitt}) and
(\ref{eq:xirr}), respectively
\begin{equation}
\label{xipm}
\xi_+(\theta)=\xi_{tt}(\theta)+\xi_{rr}(\theta),
\quad \mbox{and}\quad
\xi_-(\theta)=\xi_{tt}(\theta)-\xi_{rr}(\theta).
\end{equation}
The E and B modes' (which we denote by $M_{\perp}$) aperture mass variances
are derived by integration of these correlation functions with an
appropriate window
\begin{equation}
\label{MapVarE}
\langle M_{ap}^2 \rangle (\theta_{ap})=
\pi \int_0^{2\theta_{ap}} d \theta~\theta
\left[ {\cal{W}}(\theta)\xi_+(\theta)+
{\tilde{{\cal{W}}}}(\theta)\xi_-(\theta) 
\right],
\end{equation}
and
\begin{equation}
\label{MapVarB}
\langle M_{\perp}^2 \rangle (\theta_{ap})=
\pi \int_0^{2\theta_{ap}} d \theta~\theta
\left[ {\cal{W}}(\theta)\xi_+(\theta)-
{\tilde{{\cal{W}}}}(\theta)\xi_-(\theta) 
\right],
\end{equation}
where ${\cal{W}}$ and ${\tilde{{\cal{W}}}}$ are given in Crittenden et
al. (2002); useful analytical expressions were derived by Schneider
et al. (2002).

\vspace{0.3cm}
\centerline{{\vbox{\epsfxsize=8.0cm\epsfbox{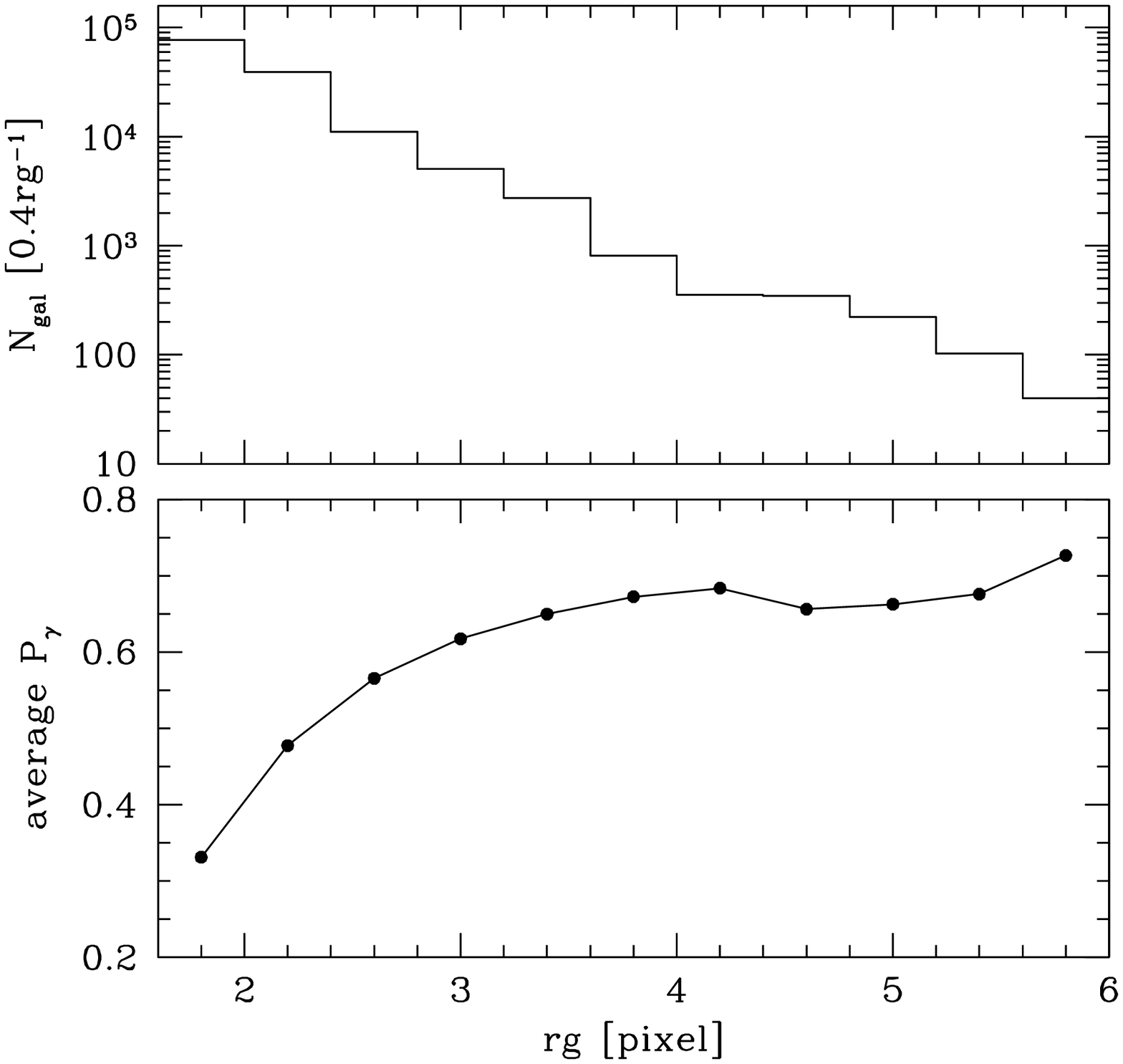}}}}
\figcaption{Top panel: the size distribution of objects used for 
the cosmic shear measurements. 
Bottom panel: Average tr$(P_\gamma)$ as a function of $rg$.
\label{fig:Pgamma}}
\vspace{0.3cm}

\centerline{{\vbox{\epsfxsize=8.0cm\epsfbox{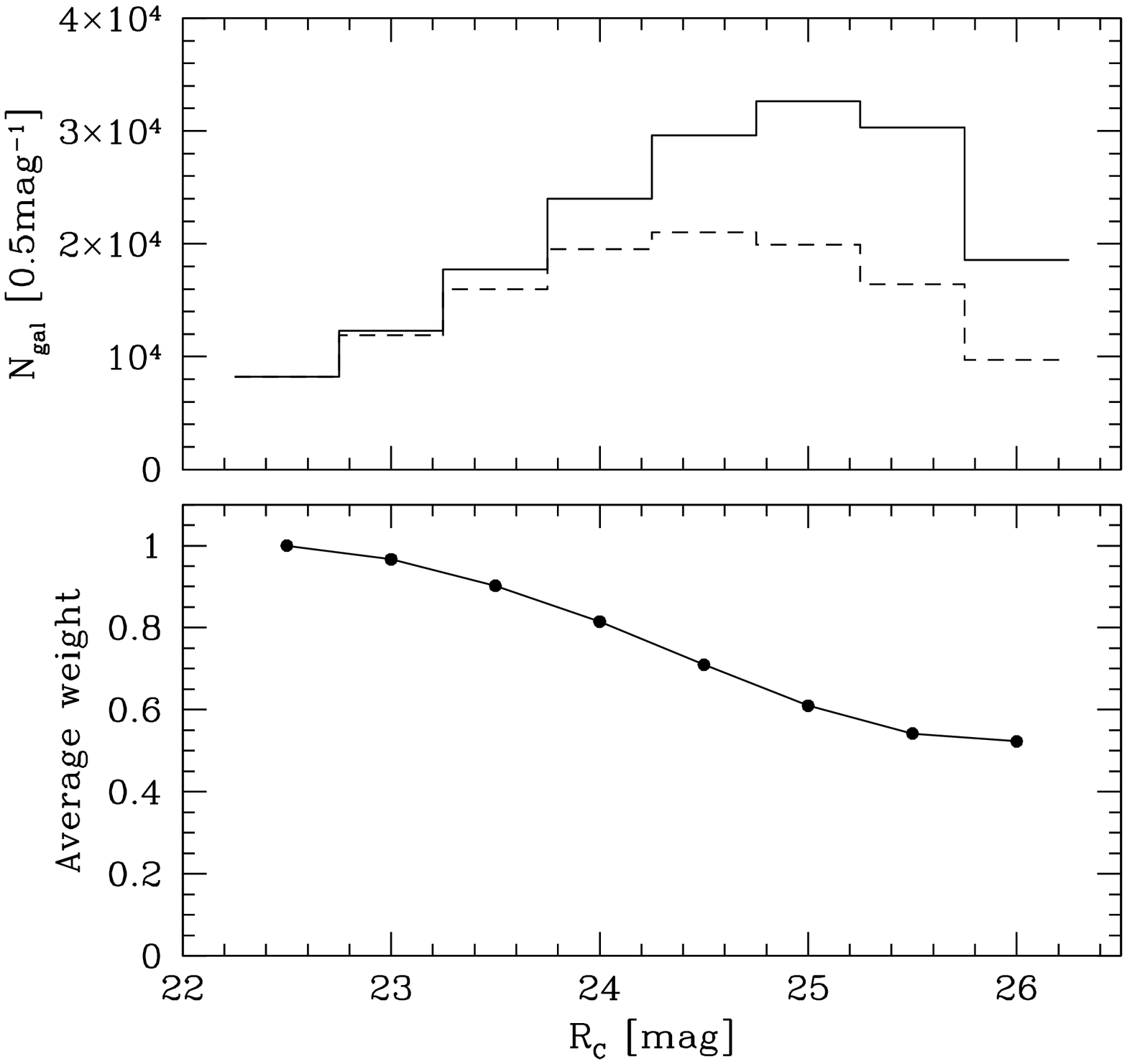}}}}
\figcaption{Top panel:the solid histogram shows the number counts of objects 
used for the cosmic shear measurements, while the dashed histogram 
is for the product of the number counts and the average weight. 
Bottom panel: Average weight as a function of $R_C$ magnitude.
\label{fig:weight}}

\section{Data}

Observations were made with the Suprime-Cam on the Subaru 8.2-m
telescope during its commissioning phase.
We used a field size of 2.1 deg$^2$, which was the largest size
possible during that period.

The field that we chose was centered at
R.A. = 16$^h$04$^m$43$^s$, decl. = +43\arcdeg12\arcmin19\arcsec
(J2000.0).
We obtained $R_c$-band images on the nights of 2001 April 23-25.
Suprime-Cam has a field of view of 34\arcmin$\times$ 27\arcmin
with a scale 0\arcsec.202 pixel$^{-1}$ (see Miyazaki et al.~2002
for instrumental details of Suprime-Cam).
Nine contiguous fields were observed in a 3$\times$3 mosaic
configuration.
Each exposure on a given field was offset by $1\sim2$\arcmin from
the other exposures
to allow the removal of cosmic rays and defects on the CCDs. The
total exposure time of each field was 1800 sec (360 sec $\times$ 5).

We apply the weak lensing mass reconstruction technique to
these data (Miyazaki et al. 2002a).
The number counts of high peaks (above 5-$\sigma$), which
represent massive dark halos, in the reconstructed convergence
field is $4.9\pm 2.3$, where the Gaussian smoothing radius of the
convergence map is $1\arcmin$.
This result is consistent with predictions thatassume the Press-Schechter mass function (Press \& Schechter 1974;
we used the version modified by Sheth \& Tormen 1999)
and the universal NFW halo profile (Navarro, Frenk \& White 1996) under
the cluster normalized CDM cosmology.
Thus, it is unlikely that this field is significantly far from the
cosmic mean.

The individual images were de-biased and then flattened using a median
of all the object frames taken during the observing run. Stacking the
dithered images is not a trivial procedure because of the large
distortion in the optics, combined with the alignment error of
each CCD with displacement and rotation from its nominal position.
We present here an outline of how to obtain parameters
that transform a CCD coordinate to a standard celestial
coordinate. First, we employ a geometrical model of the field
distortion using a fourth order polynomial function
\begin{equation}
\frac{R - r}{r} = ar + br^2 + cr^3 + dr^4,
\end{equation}
where $R$ and $r$ are the distance from the optical axis in units of 
pixels on the face of the CCD and in celestial coordinates, respectively.
We typically changed the telescope pointing by 1$\sim$2
arcmin between successive exposures to fill the gaps between the CCDs.
The offset and rotation of the telescope pointing between
the exposures are also set as free parameters. All of these parameters
can be determined by minimizing the distance of the same stars
identified on different exposures.
To do this, we adopt a modified version of
$mosaicfit$, which is one of the functions of the $imcat$ suite.

Once we have obtained these parameters, we use them to warp each image
before stacking. The residual of the distances between corresponding
star images is a measure of the error of this mosaic-stacking procedure.
The RMS value of the residuals is typically about 0.5 pixels.
As shown in Miyazaki et al. (2002b), distortion parameters obtained in
this way match quite well with the residuals predicted by optical
ray-tracing programs, which implies that our solution is satisfactory.

The RMS residual of 0.5 pixels is due to several effects that are not
considered in the simple model, including atmospheric dispersion
and asymmetric aberration of the optics. We note, however, that the
residual vector changes smoothly with position and can be well
modeled as a third order bi-linear polynomial function of position.
This model then gives a fine correction to our geometrical solution
described above.
The measurements of the displacement and the warping of the images
are carried out using {\it fitgeometry2} and $mosaicmap$ of $imcat$,
respectively, as described by Kaiser et al. (1999).
As a result of these procedures, the final residual typically decreases
down to 0.07 pixel RMS (14 milliarcsec).
To derive a better astrometric solution, an external
reference star catalog would have to be used.
However, we simply employ each first exposure as a
reference and accept the resulting moderate astrometric accuracy,
as it does not significantly affect the weak lensing analysis.
The individual images are then warped using the solution,
and stacked.
The seeing in the resulting image is 0\arcsec.68 FWHM and the scatter
among the fields is quite small at 0\arcsec.04 rms.

To carry out object detection, photometry and shape measurements of
objects, we use {\it hfindpeaks}, {\it apphot} and {\it getshapes}
of the $imcat$ suite, which are an implementation of Kaiser
et al.~(1995).
Catalogs created for the nine fields are registered using stars in the
overlapping regions to generate a final catalog whose total field of
view is 1.64\arcdeg $\times$ 1.28\arcdeg.
Differences in the photometric zero point among the fields due to
variation in the sky conditions (in turn, due to thin cirrus and
differing air mass) are compensated for at this stage, but the
adjustment is not significant ($\sim$ 0.05 mag).

We adopt slightly different object selection criteria to
those used by Miyazaki et al. (2002a) to optimize our cosmic shear
correlation measurement.
Our criteria are
(i) $22.5 < R_c < 26$.
(ii) The signal-to-noise (S/N) ratio, $nu$, calculated in {\it
imcat}, exceeds 7 (Erben et al. 2001).
(iii) The image size is larger than the PSF size, $rg>1.45\sim 1.65$,
where $rg$ is a measure of the size of objects (in pixel units) yielded
by {\it imcat}.
The PSF size is identified from the stellar branch in the size-magnitude
($rg$-$R_c$) plane.
The PSF size varies slightly from pointing to pointing because of
changes in the seeing conditions.
(iv) Objects with $rh>10$ pixels (where $rh$ is the half light
radius) are considered too large and removed.
(v) Highly elliptical objects, where the observed ellipticity,
$|e^{obs}|>0.5$, are removed.
(vi) Objects that have a close companion, with a separation of less
than 10 pixels ($\simeq2$ arcsec), are removed to avoid the
problem of overlapping isophotes reported by Van Waerbeke et al. (2000).
The number of objects that pass these various selection criteria (i)-(vi) is
249,071 (32.9 arcmin$^{-2}$).

Van Waerbeke et al.~(2000) reported that regions where
data from different CCDs are stacked together as a result of
offsets between exposures (specifically, the edges of CCDs)
can potentially produce discontinuities in the properties of the
field and thus make the PSF correction difficult.
This effect could cause a systematic error in the cosmic shear
correlation measurement. Therefore, we decided to mask such regions.
That is, we use only objects that were observed by the same CCD.
About 35\% of the objects are removed by this masking, and
the number of objects in the final catalog is 161,740.
The image size distribution and magnitude distribution of the catalog
are shown in the upper panel of Figure \ref{fig:Pgamma} and Figure
\ref{fig:weight}, respectively.

\section{Galaxy shape analysis}

The observed ellipticity of galaxies is measured from the weighted
quadrupole moments $I_{ij}$ of the surface brightness $f(\bm{\theta})$:
\begin{equation}
\label{eobs}
\bm{e_{obs}}=\left({{I_{11} - I_{22}}\over {I_{11}+I_{22}}},
{{I_{12}}\over {I_{11}+I_{22}}} \right),
\quad
I_{ij}=\int d^2\theta ~ W_G(\theta) \theta_i \theta_j f(\bm{\theta}),
\end{equation}
where $W_G(\theta)$ is the Gaussian window function.
Estimating the shear, $\bm{\gamma}$, from the observed
ellipticities, $\bm{e_{obs}}$, involves two steps:
First, the point spread function (PSF) anisotropy is corrected using
the star images as references,
\begin{equation}
\label{Psmcorrection}
\bm{e} = \bm{e_{obs}} - \frac{P_{sm}}{P_{sm}^{*}}\bm{e_{obs}^*},
\end{equation}
where $P_{sm}$ is the smear polarisability tensor, which is
mostly diagonal (Kaiser et al. 1995).
$(P_{sm}^{*})^{-1}\bm{e_{obs}^{*}}$ was calculated for stars scattered
over the field of view.
We use unsaturated stars selected by the following criteria:
(i) $20.6 < R_c < 23.0$ (using fainter stars than this does not change
the results, see the Appendix). Note that the saturation level identified
from the size-magnitude ($rg$-$R_c$) plane is $R_c = 19.5\sim 20.5$
which correlates with the seeing (with a fainter saturation level for
a better seeing).
(ii) $nu>15$.
(iii) The image size is within a narrow range of the seeing size,
$rg_*-0.05<rg<rg_*+0.05$, where $rg_*$ denotes the central $rg$ value
(in pixel units) of the stellar branch in the $rg$-$R_c$ plane.
$rg_*$ varies from pointing to pointing and is $rg_*=1.27\sim1.52$.
(iv) Highly elliptical objects, with an observed ellipticity, $|e|>0.3$,
are removed.
As a result, the average number density of the selected stars is about
1/arcmin$^2$, and there are on average about 60 stars in each chip.
However, the chip-by-chip variation in the number of stars is quite
large; some chips have only $\sim$20 stars because of the presence of
a large saturated star in the field.
We make the first order bi-polynomial fit to values of
$(P_{sm}^{*})^{-1}\bm{e_{obs}^{*}}$ as a function of position
(a second order fit does not change the results, see the Appendix).
To make this fit, we use the {\it efit} command in {\it imcat}, and
flux weighting is not applied.
This function is used in eq.~(\ref{Psmcorrection}) to correct the
ellipticities of faint galaxies.
This correction is made with the {\it ecorrect} command in {\it imcat}.
Note that each CCD chip is treated independently in this procedure.
We found that a first order fit corrects the PSF anisotropy well,
and furthermore does not introduce a systematic artificial residual
due to the wings of a higher order fit (Van Waerbeke et al.~2002),
provided that overlapped CCD regions are masked.
The RMS value of the ellipticities of the reference stars, $\langle
|\bm{e^{*}}|^2 \rangle ^{1/2}$, is reduced from 2.8\% to
1.0\% as a result of the correction. Note that the RMS before
the correction is already small, because of the superb image quality of
the Subaru telescope.
Figure \ref{fig:eep} shows the star ellipticities before (left panel) and
after (right panel) the anisotropic PSF correction from one pointing.
The observed ellipticities not only have a large scatter but also show
a systematic effect. After the PSF correction this tendency is
removed and the corrected star ellipticities are distributed
symmetrically around zero with a small scatter.

\vspace{0.3cm}
\centerline{{\vbox{\epsfxsize=8.4cm\epsfbox{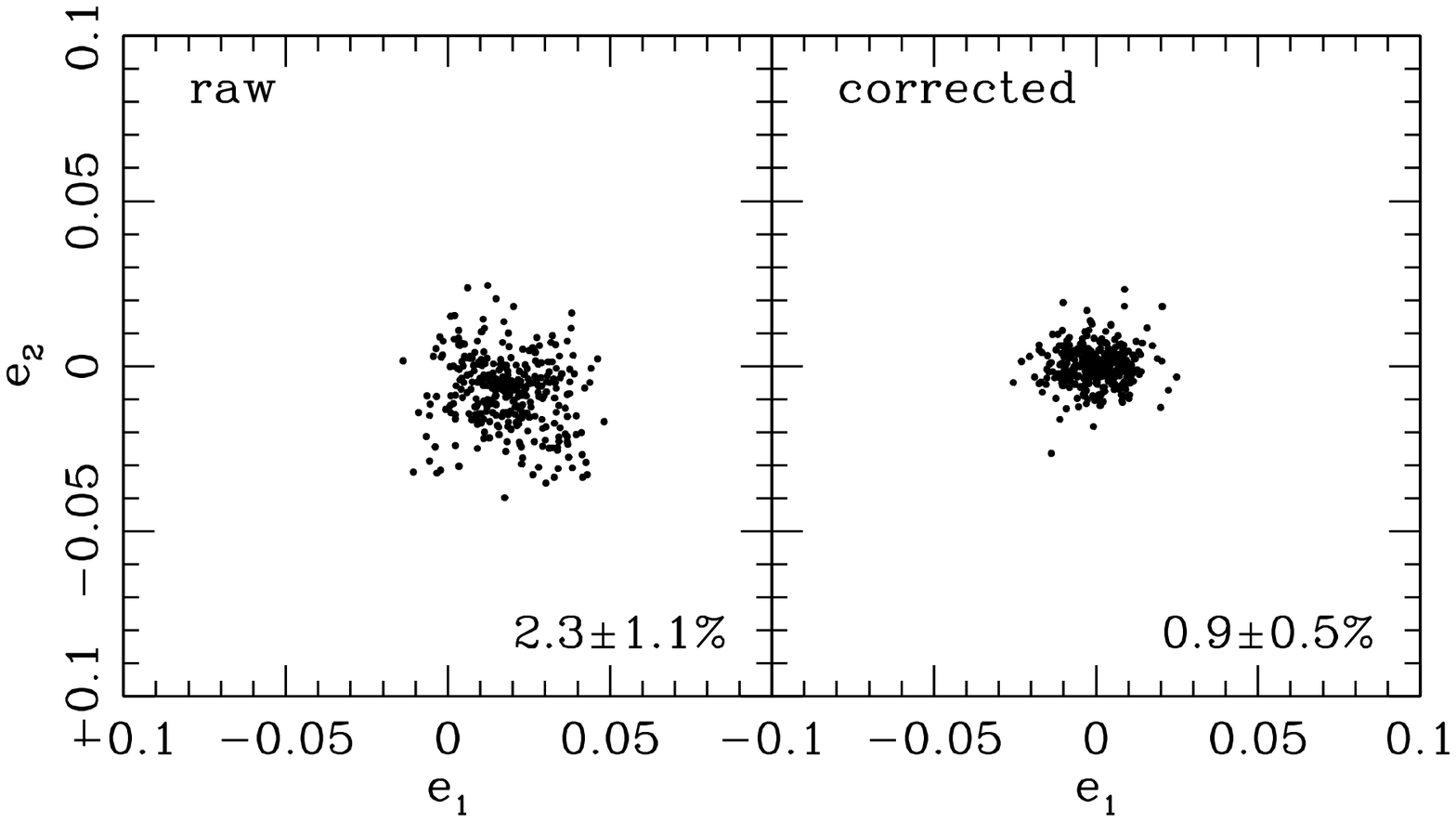}}}}
\figcaption{Ellipticities distribution of stars, before (left) and after 
(right) the correction for PSF anisotropies is made. 
The mean and dispersion of the ellipticites ($|\bm{e}|$) of all the stars are 
shown inside the frame.
\label{fig:eep}}

\centerline{{\vbox{\epsfxsize=8.4cm\epsfbox{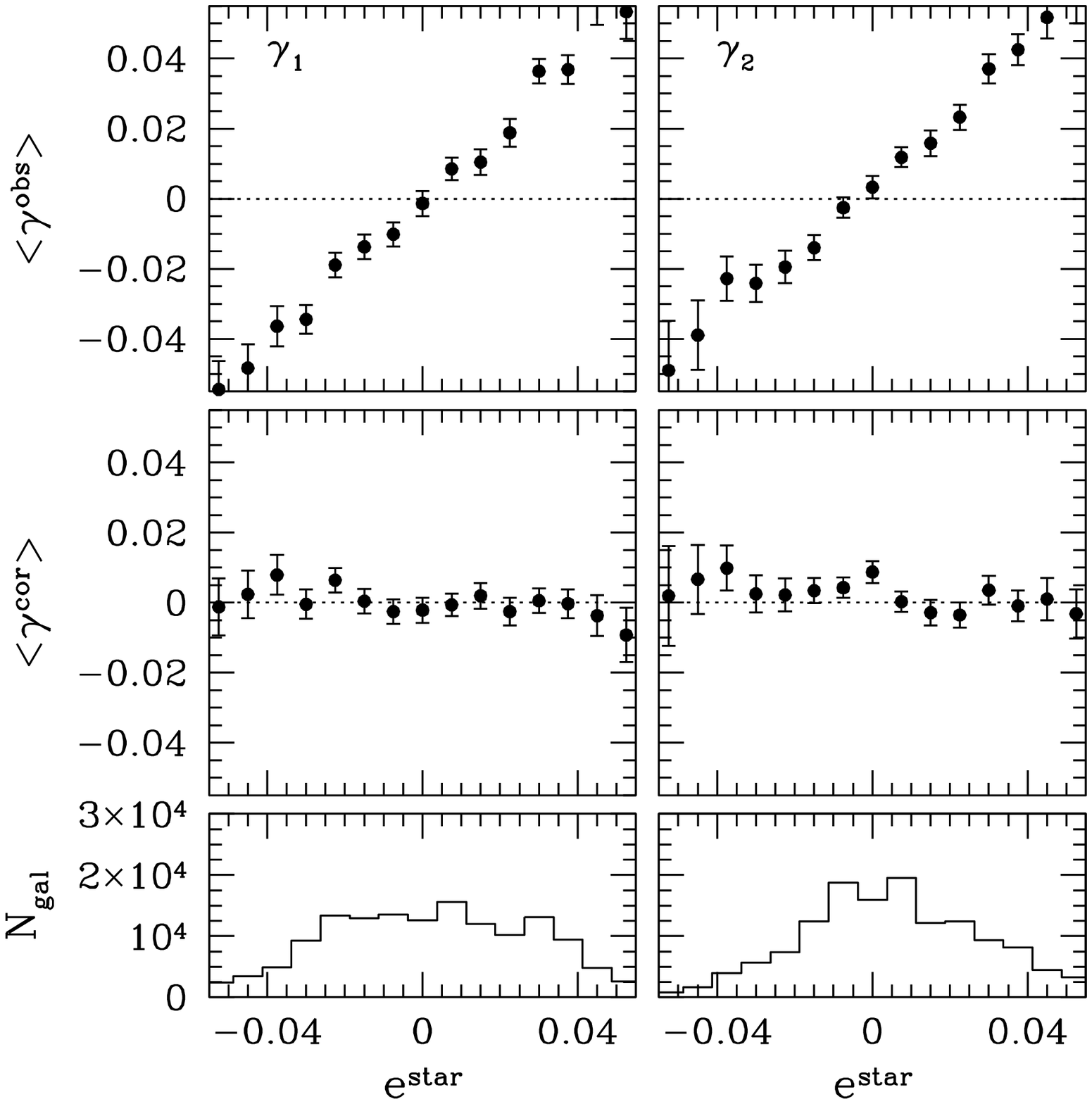}}}}
\figcaption{Components of galaxy ellipticities as a function of the 
star ellipticity component used for the anisotropic PSF correction. 
Top and middle panels show the averaged 
galaxy ellipticities before and after correction, respectively.
Bottom panels show the ellipticity distribution of galaxies (number
per bin) used  for the cosmic shear correlation measurement.
\label{fig:egal_estar}}
\vspace{0.3cm}

Figure \ref{fig:egal_estar} shows the average ellipticity of galaxies
as a function of the value of the star ellipticity used for the
anisotropic PSF correction.
The top panels show the observed ellipticities, and the expected
strong correlation is present.
Note that for the majority of galaxies, the PSF anisotropy is
very small, as shown in the bottom panels.
The corrected ellipticities plotted in the middle panels comprise
only small values.
The averaged values of the corrected ellipticities are
$\langle e_1 \rangle=-5.0\times 10^{-4}$ and $\langle e_2
\rangle=1.8\times 10^{-3}$. Thus, no significant offset is found.

The second step is the isotropic correction, caused by the window, $W_G$,
and the seeing.
Luppino \& Kaiser (1997) reported a method to correct the
ellipticities for these effects.
The {\it pre-}seeing shear $\bm{\gamma}$ is described as
\begin{equation}
\label{eq:Pgamma}
\bm{\gamma} = P_{\gamma}^{-1} \bm{e}, \quad
P_{\gamma} = P_{sh} - \frac{P_{sh}^{*}}{P_{sm}^{*}}P_{sm},
\end{equation}
where $P_{sh}$ is the shear polarisability tensor. The $P_{\gamma}$ of
individual galaxies is, however, known to be a noisy estimate, and
thus we adopt the smoothing and weighting method developed by
Van Waerbeke et al.(2000; see also Erben et al. 2001 for a detailed
study of the smoothing scheme).
For each object, 20 neighbors are first identified in the
$r_g$-$R_c$ plane.
A median value of $P_{\gamma}$ among these neighbors is adopted as the
smoothed $P_{\gamma}$ of the object.
The averaged ${\rm tr}(P_{\gamma})$ for all objects is calculated as
$0.40$ but $P_{\gamma}$ depends on the object size. 
Figure \ref{fig:Pgamma} plots the averaged tr$(P_{\gamma})$ as a
function of $rg$.
This graph shows that the average ${\rm tr}(P_{\gamma})$
is almost constant, $\sim 0.65$ for $r_g > 3.5$, but becomes
smaller for smaller $r_g$.

As the estimated $\bm{\gamma}$ is still noisy, especially for small
and faint objects, it is important to weight the galaxies according to
the uncertainty in the shape measurements.
We weight the objects using the procedure developed by
Van Waerbeke et al.~(2000; 2002; see also Erben et al. 2001).
The variance of raw $\bm{\gamma}$ before the smoothing among the
neighbors, $\sigma_\gamma^2$, is used to estimate the
weight of each object, $w$, as
\begin{equation}
\label{eq:weight}
w = {1\over{\sigma_\gamma^2 + \alpha^2}}
\end{equation}
where $\alpha^2$ is the variance of all the objects in the catalog
and $\alpha\simeq 0.4$ here.
Under the weighting scheme, an averaged value of a certain observable,
$\langle A \rangle$, is calculated as $\sum^N_{i=1}w_{i}A_i /
\sum^N_{i=1}w_{i}$ instead of $\sum^N_{i=1}A_i/N$.
The lower panel of Figure \ref{fig:weight} shows the averaged weight
as a function of $R_c$ magnitude.
As expected, less weight is given to fainter objects because
the shape measurements of these objects are noisier.
The dashed histogram in the upper panel of Figure \ref{fig:weight}
plots the product of the number counts and the average weight
and shows that the {\it effective} counts peak at $R_c\sim 24.5$.

\vspace{0.3cm}
\centerline{{\vbox{\epsfxsize=8.4cm\epsfbox{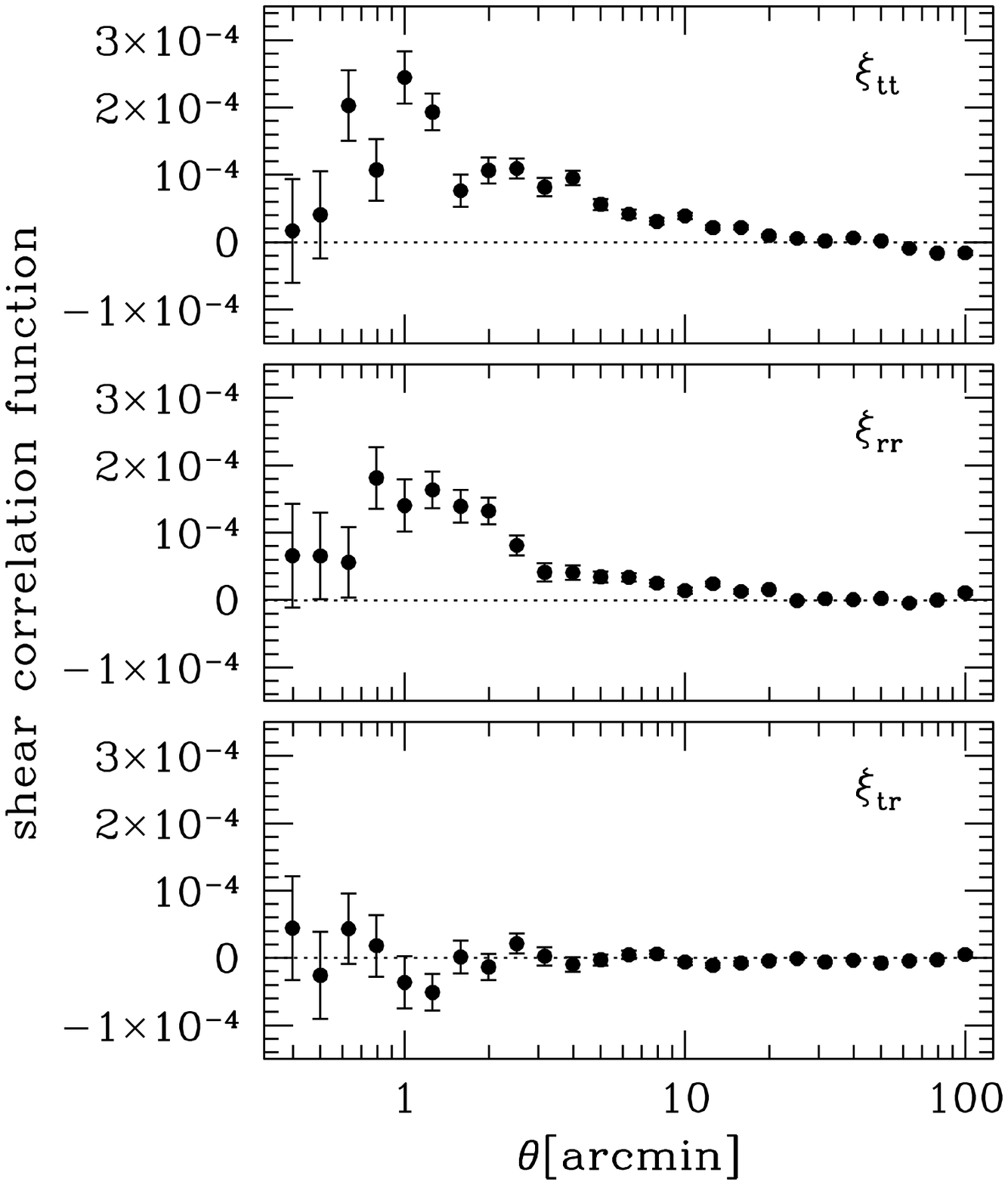}}}}
\figcaption{Shear correlation functions. Top panel is for $\xi_{tt}$, 
middle panel for $\xi_{rr}$, and bottom for the cross correlation 
$\xi_{tr}$ which should vanish if the data are not contaminated 
by systematics. The error bars present 
the statistical error computed from 100 randomized realizations.
\label{fig:xi3}}

\section{Cosmic shear correlations}

In this section, we present the cosmic shear correlations measured
from the 2.1 deg$^2$ Suprime-Cam data, and discuss their statistical
and possible systematic errors.

The top and middle panels of Figure \ref{fig:xi3} show the shear
correlation
functions, $\xi_{tt}$ and $\xi_{rr}$, respectively, computed using the
estimator eqs.~(\ref{eq:xitt}) and (\ref{eq:xirr}).
The bottom panel presents the cross-correlation $\xi_{tr}$.
The error bars indicate the RMS among 100
randomized realizations, in which the orientations of galaxies are
randomized, and presumably represent the statistical error.
In the following, statistical errors are computed in this manner.
As the two top panels clearly show, we detect non-zero shear correlation
signals on scales below 30 arcmin, except for the smallest two bins.
On small scales ($\theta<1$ arcmin) the number of pairs corresponding
to the separations decreases, and consequently the signals become noisy
and the statistical errors become large.
As the cross-correlation should be zero for a signal due to
gravitational lensing, it provides a check on systematic effects in the
data.
The bottom panel shows that the cross-correlation is indeed consistent
with zero at all scales.
This result indicates that our PSF corrections perform well and
do not introduce a systematic effect into the data.

\vspace{0.3cm}
\centerline{{\vbox{\epsfxsize=8.4cm\epsfbox{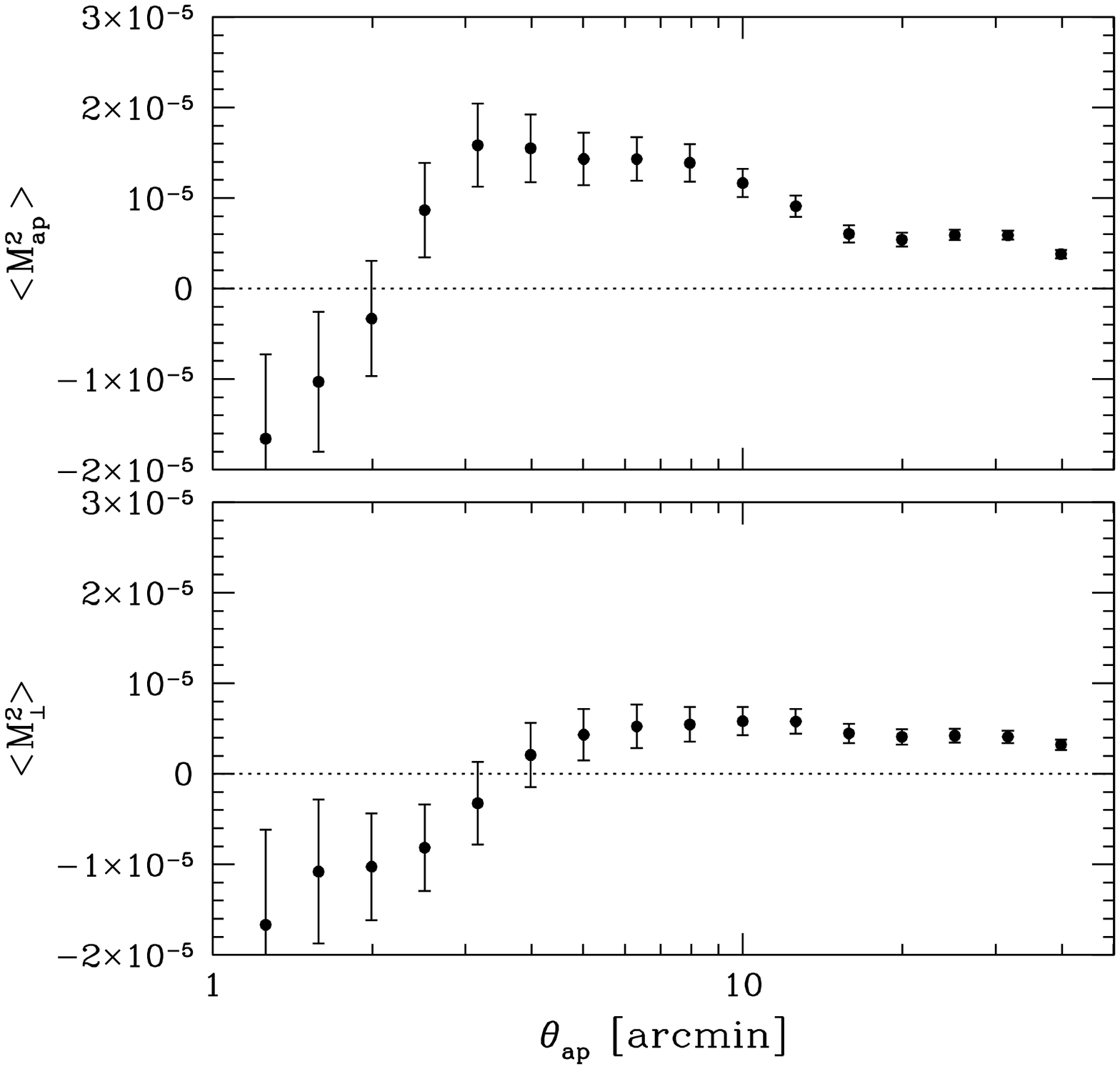}}}}
\figcaption{Upper panel shows the E-mode aperture mass variance 
$\langle M_{ap}^2 \rangle$, while the lower panel for B-mode aperture 
mass variance $\langle M_{\perp}^2 \rangle$. The error bars present 
the statistical error computed from 100 randomized realizations.
\label{fig:MapEB}}
\vspace{0.3cm}

Let us now turn to the aperture mass variance.
We compute the E and B mode aperture mass variances from the two-point
shear correlation functions via eqs.~(\ref{MapVarE})
and (\ref{MapVarB}).
We use the analytic expressions for the window
function ${\cal{W}}$  and ${\tilde{{\cal{W}}}}$ given in Schneider et
al.~(2002).
The shear correlation functions defined by eq.~(\ref{xipm}) are computed
over the range $0.04\arcmin<\theta<90\arcmin$ on 168 bins equally spaced
with a log-interval of $\Delta \log \theta=0.02$.
The E and B mode aperture mass variances are plotted in Figure
\ref{fig:MapEB}.
For the E mode, we detect positive, non-zero signals on scales larger
than $\theta_{ap}>2\arcmin$. Since, as shown in Figure \ref{fig:xi3},
the two-point correlation function becomes very noisy on scales smaller
than 1 arcmin, and also considering that the aperture mass
effectively probes a real scale of $\sim\theta_{ap}/5$,
we use only the signals on scales larger than
$\theta_{ap}=2$ arcmin for our maximum likelihood analysis in the next
section.
The amplitude and main features of the E mode variance are broadly
consistent with theoretical predictions under the cluster normalized
CDM cosmology (see Figure \ref{fig:map_error}).

The B mode aperture mass variance is shown in the lower panel of
Figure \ref{fig:MapEB}.
Small but non-zero signals are found on scales larger than 5 arcmin.
Currently, the origin of this B mode variance is not clear.
One possibility is incorrect anisotropic PSF correction.
We tested this possibility by repeating the PSF correction using
different procedures, and found no significant problems (see the
Appendix).
It is interesting to note that Van Waerbeke et al. (2002) reported
the detection in their data of a non-zero B mode variance
($\langle M_{\perp}^2\rangle\sim3\times10^{-6}$) on scales
$10\arcmin<\theta_{ap}<40\arcmin$, the amplitude of which is very
similar to ours.
Note that their survey depth (a limiting magnitude of $I_{AB}=24.5$)
was slightly shallower than ours.
On the other hand, Hoekstra et al.~(2002b) reported a vanishing B mode
variance on scales larger than 10 arcmin in their shallow data
($R_c<24$).
These results suggest that the current procedures for galaxy
shape corrections become problematic for fainter objects.
Unfortunately, we cannot test this possibility because the number of
brighter galaxies is not sufficient to obtain meaningful statistics.
This possibility should be tested in future studies.

Currently, it is not clear how to correct the E mode variance,
given the observed B mode.
We follow Van Waerbeke et al.~(2002b) and add the B mode in quadrature
to the uncertainty in the E mode for the maximum likelihood analysis
(see \S 6).

\vspace{0.3cm}
\centerline{{\vbox{\epsfxsize=8.4cm\epsfbox{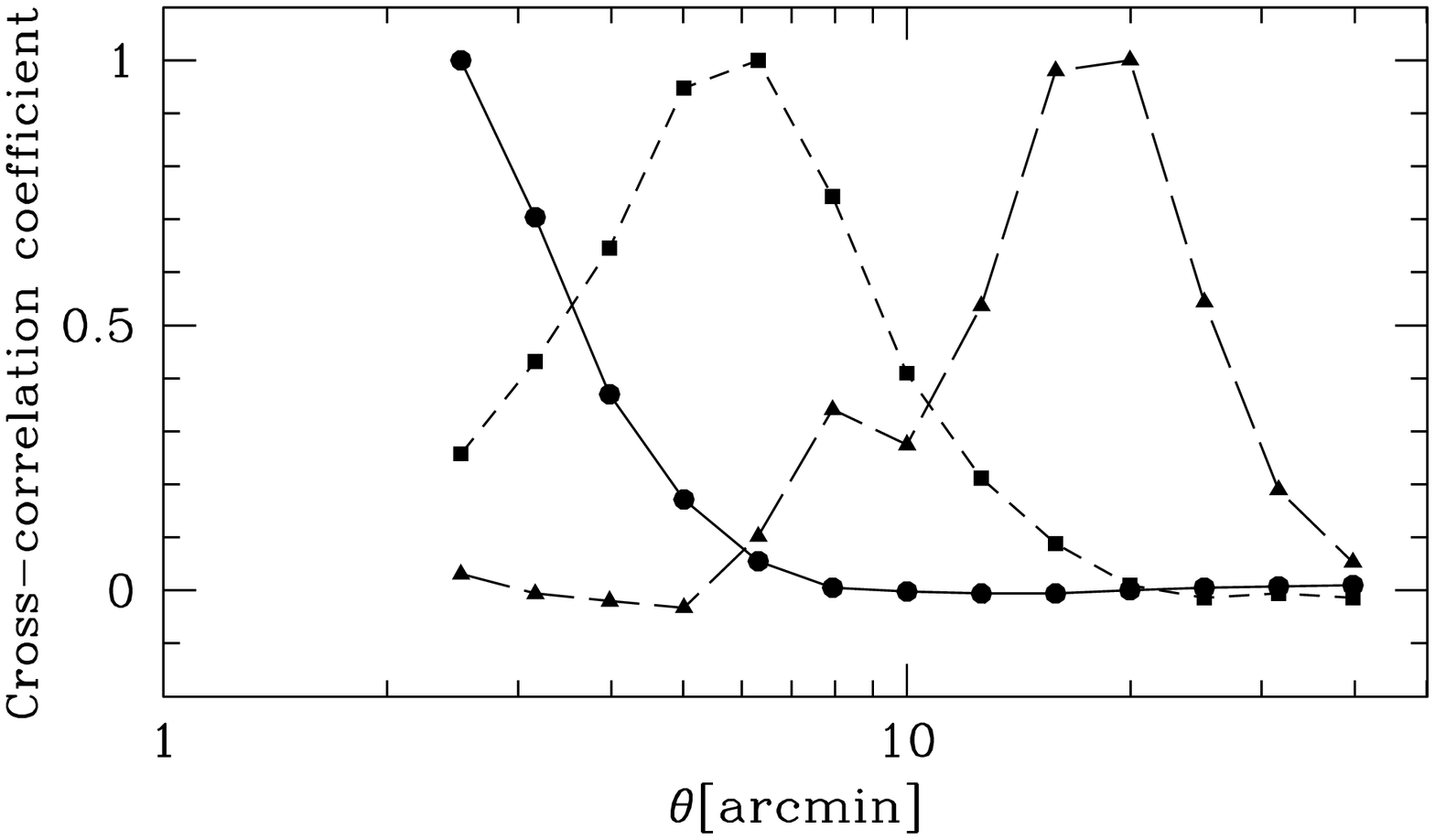}}}}
\figcaption{Cross-correlation coefficient $r(\theta,\theta')$ for the 
statistical noise as a function of scale $\theta$ for three scales, 
$\theta'=2.5$ (filled circles with solid line), 6.3 (filled squares with
dashed line)and 20 arcmin (filled triangles with long-dashed line).
\label{fig:cross}}

\section{Likelihood analysis}

In this section, we compare the measured aperture mass variance
to the model predictions in CDM cosmologies using maximum likelihood
analysis. We present constraints on cosmological parameters
obtained from the cosmic shear variance observed in the Suprime-Cam
2.1 degree$^2$ field data.

\subsection{Source redshift distribution}
To compute the theoretical prediction of the aperture mass
variance using eq.~(\ref{eq:Mapvar}), we need to fix the redshift
distribution of the source galaxies.
However, no redshift information about the galaxies in our catalog is
available.
Therefore, we decided to adopt a parameterized model that provides
a good fit to the redshift distribution of deep surveys
(e.g., Hoekstra et al. 2002b; Van Waerbeke et al. 2002),
\begin{equation}
\label{eq:zdist}
n_s(z)={{\beta}\over{z_* \Gamma\left[(1+\alpha)/\beta\right]}}
\left(z\over{z_*} \right)^\alpha
\exp\left[- \left( z\over{z_*}\right)^{\beta}\right],
\end{equation}
with $\alpha=2$ and $\beta=1.5$.
The mean redshift relates to
$z_*$ as $\bar{z}_s = z_*\Gamma\left[(2+\alpha)/\beta\right]/
\Gamma\left[(1+\alpha)/\beta\right]$ and for these values of
$\alpha$ and $\beta$, it gives $\bar{z}_s \simeq 1.5 z_*$.
The median redshift is approximately given by $z_{med} \simeq 1.4 z_*$.
We treat the mean redshift of the distribution as a model parameter
in the maximum likelihood analysis.

\subsection{Maximum likelihood analysis}

In performing the maximum likelihood analysis, we basically follow the
procedure described in Van Waerbeke et al. (2002; see also Hoekstra
et al.~2002b).
The theoretical predictions are computed in a four-dimensional space,
but we restrict the parameter space to realistic but conservative
ranges:
$\Omega_{\rm m}\in[0.1,1]$ (either $\Omega_\Lambda=0$ or
$\Omega_{\rm m}+\Omega_\Lambda=1$),
$\sigma_8\in[0.2,2]$,
$\Gamma\in[0.05,0.75]$ and
$\bar{z}_s \in[0.3,2.5]$ with a sampling of
$19 \times 19 \times 11 \times 23$.
In what follows, we refer to this parameter range
($\Omega_{\rm m}$, $\sigma_8$, $\Gamma$, $\bar{z}_s$)
as the {\it default prior} space.
The model predictions are then interpolated with an oversampling 
fivefold higher in each dimension.

\centerline{{\vbox{\epsfxsize=8.4cm\epsfbox{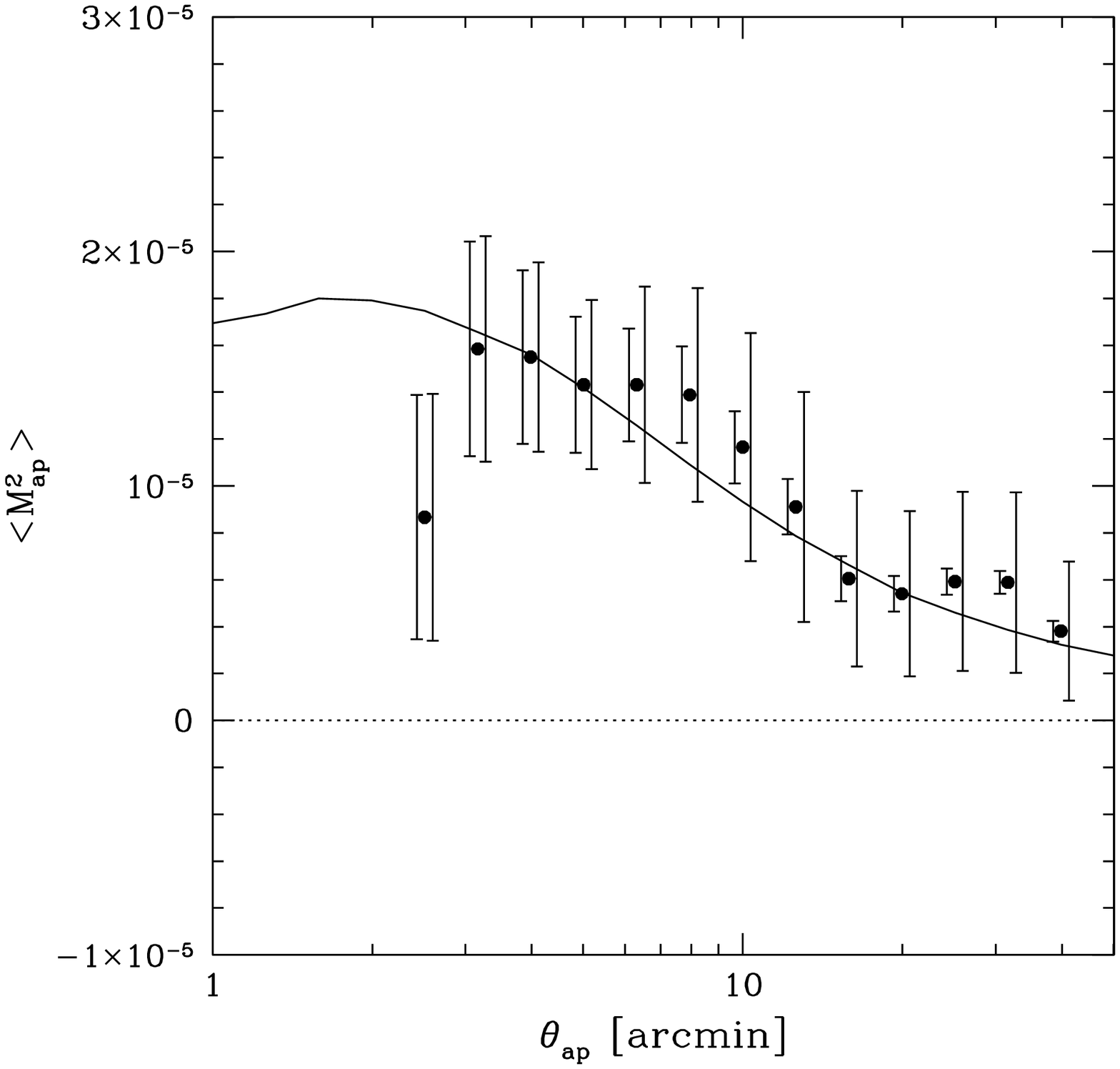}}}}
\figcaption{The aperture mass variance. For each measurement point, 
the left error bars show statistical error while right error bars present
a sum of the statistical error, a residual systematic error 
estimated from B mode variance and the cosmic variance in quadrature.
On smaller scales, statistical error dominates, while on larger scales 
residual error dominates. 
Comparing with them, the cosmic variance does not have a serious
impact on the final error on all scales.
The solid curve shows, for an illustrative example, the theoretical 
prediction of CDM model with $\Omega_{\rm m}=0.3$, 
$\Omega_{\Lambda}=0.7$, $\Gamma=0.21$, $\sigma_8=0.85$ and $\bar{z}_s=1$.
\label{fig:map_error}}
\vspace{0.3cm}

For a given theoretical model, we compute the $\chi^2$ (log-likelihood)
\begin{equation}
\label{eq:chi2}
\chi^2 = (d_i-m_i)\mbox{\bf C}^{-1}(d_i-m_i)^T,
\end{equation}
where $d_i$ is the measurement at scale $\theta_i$, and $m_i$ is the
corresponding theoretical prediction.
Confidence values are computed from this $\chi^2$ in the standard
manner.
The covariance matrix consists of three contributions
\begin{equation}
\label{eq:cov}
C(\theta_i,\theta_j)=C_s(\theta_i,\theta_j)+C_b(\theta_i,\theta_j)+
C_{cv}(\theta_i,\theta_j),
\end{equation}
where $C_s$,  $C_b$ and $C_{cv}$ are the statistical error, the residual
systematics, and the cosmic variance, respectively, computed by the
procedures described below.
The statistical error was computed from the 100 randomized realizations
catalog.
Figure \ref{fig:cross} shows the cross-correlation coefficient for
the statistical noise for three scales, 2.5, 6.3, and 20 arcmin,
with other scales.
The non-zero B mode variance could indicate the existence of a residual
systematic, although its origin is not yet clear.
It would be natural to consider that a residual systematic of similar
size also exists in the E mode.
However, there is not yet a clear scheme to deal with this
residual systematic.
Therefore, we adopt the simple and conservative procedure
described in Van Waerbeke (2002).
We quadratically add the B mode variance to the error of the signal.
As the E and B mode covariance matrices for the statistical
noise are identical, the diagonal part of the matrix $C_b$ is given
by the B mode signal and off-diagonal terms
follow the same correlation properties as the E mode.

Estimation of the cosmic variance is more complicated, because
the observed scales are in the quasi-linear to nonlinear regime
and thus random Gaussian theory cannot be applied.
The cosmic variance is estimated using weak lensing numerical
experiments, which are performed using a ray-tracing technique
combined with large $N$-body simulations; the details are described
in Hamana et al.~(in preparation; see also Menard et al.~2003; 
Takada \& Hamana 2003).
Briefly, the dark matter distribution from the observer ($z=0$) to
high $z$ ($z\sim 3$) is generated by stacking 10 snapshot outputs from
the $N$-body simulations.  $N$-body data from the Very Large Simulation
(VLS) which followed $512^3$ particles in a cubic box of 479$h{^1}$Mpc
on a side, carried out by the Virgo Consortium (Jenkins et al. 2001;
Yoshida, Sheth \& Diaferio 2001) are used.
A $\Lambda$CDM model ($\Omega_m=0.3$,
$\Omega_\lambda=0.7$ and $h=0.7$) is assumed with the CDM initial
power spectrum computed using CMBFAST (Seljak \& Zaldarriaga 1996).
The multiple-lens plane ray-tracing algorithm is used to follow the
light rays (Hamana \& Mellier 2001, see also Jain, Seljak \& White
2000 for the basic theory).
The lensing convergence and shear is computed for 1024$^2$ pixels
with a pixel size of 0.25 arcmin for a single source plane of $z_s=1$.
We compute the covariance matrix due to the cosmic variance from 36
random mock observations (but without the intrinsic ellipticity of
galaxy images) generated by the numerical experiment.
As the ratio of the non-Gaussian to Gaussian contributions to cosmic
variance does not vary significantly with the underlying cosmology
(Van Waerbeke et al. 2002), the cosmic variance
also does not play an important role at all scales; given the large residual
error from the B mode variance at large scales, we use the covariance
matrix obtained from the $\Lambda$CDM numerical experiment regardless
of the cosmological model considered.
Figure \ref{fig:map_error} shows the aperture mass variance with
error bars from the statistical error only (the left error bar on each
point) and a sum of the statistical, systematic and cosmic
variance in quadrature (the right error bar).
On small scales, statistical error dominates [the statistical error (left)
is comparable to the total error (right)], while at large scales
the systematic error dominates given the large B mode variance.

\vspace{0.3cm}
\centerline{{\vbox{\epsfxsize=8.4cm\epsfbox{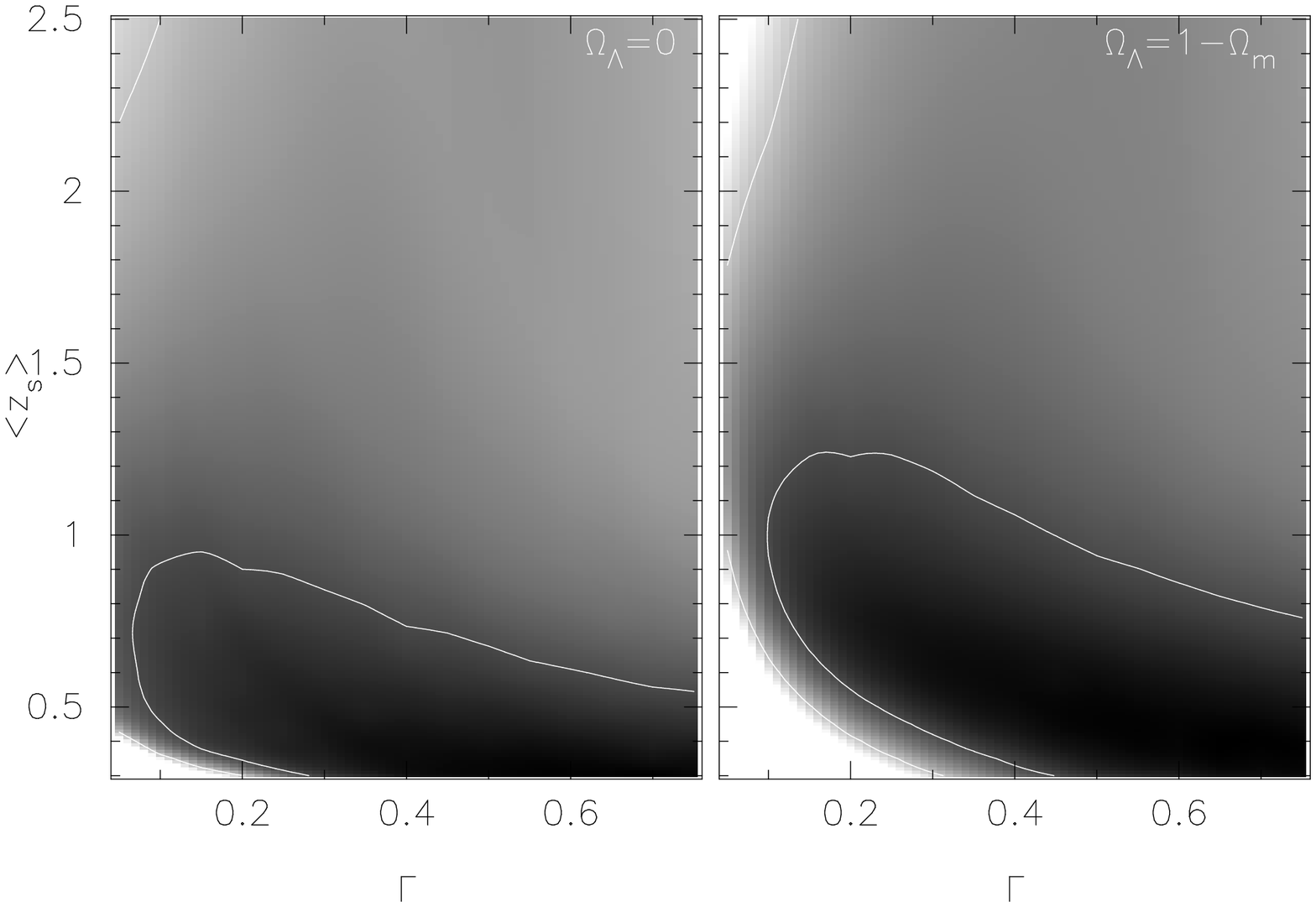}}}}
\figcaption{The gray-scale shows the $\Delta \chi^2$ map on $\Gamma$ and 
$\bar{z}_s$ obtained after a marginalization over 
$\Omega_{\rm m}\in[0.1,1]$ and $\sigma_8\in[0.2,2]$.
A darker gray-scale indicates a lower $\Delta \chi^2$ value 
(thus more likely).
Contours indicate 68.3 and 95.4\% confidence levels.
\label{fig:gz_map}}

\subsection{Results and discussion}

Here, we present two-parameter space constraints obtained
from marginalizations over the two remaining parameters.
Figure \ref{fig:gz_map} shows the likelihood map on the
$\Gamma$-$\bar{z}_s$ plane where the default prior is applied for
$\Omega_{\rm m}$ and $\sigma_8$.
This Figure clearly presents the correlation between $\Gamma$ and
$\bar{z}_s$; that is, a flatter (steeper) power spectrum
(a larger (smaller) $\Gamma$) prefers a lower (higher) $\bar{z}_s$,
(this was reported previously by Van Waerbeke et al. 2002).
The message here is that, given a relatively narrow range of
signal detection ($2\arcmin<\theta_{ap}<40\arcmin$) with large
error bars (as shown in Figure \ref{fig:map_error}), the slope of the
power spectrum and its normalization are degenerate.
Similar considerations apply to a constraint on $\sigma_8$-$\Gamma$,
as shown in the lower left panel of Figure \ref{fig:degeneracy}.
Although the constraint on $\Gamma$ and $\bar{z}_s$ is not as
tight, we can see from Figure \ref{fig:gz_map} that for
$\bar{z}_s>0.9$ ($>0.7$) a flatter power spectrum with $\Gamma>0.5$,
would be ruled out at more than the 68\% confidence level for the flat
(open) model.

\vspace{0.3cm}
\centerline{{\vbox{\epsfxsize=8.4cm\epsfbox{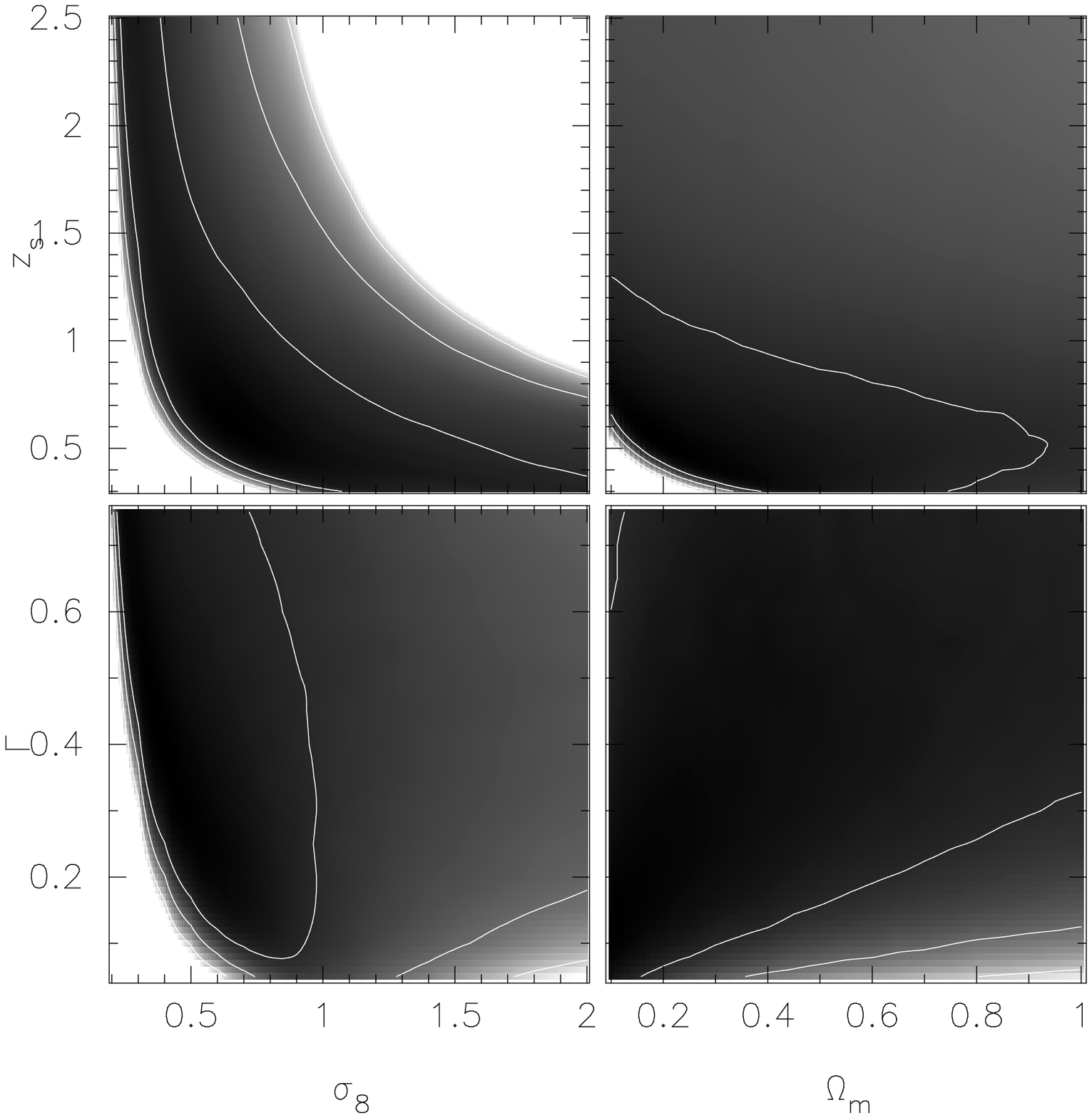}}}}
\figcaption{The gray-scale shows the $\Delta \chi^2$ on 
two-parameter space obtained after a marginalization over 
two remaining parameters for the default prior, 
a darker gray-scale for a lower $\Delta \chi^2$ value (thus more likely).
Plots are for $\sigma_8$-$\bar{z}_s$ (upper left),
$\Omega_{\rm m}$-$\bar{z}_s$ (upper right),
$\sigma_8$-$\Gamma$ (lower left) and
$\Omega_{\rm m}$-$\Gamma$ (lower right).
Contours indicate 68.3, 95.4 and 99.73\% confidence levels.
\label{fig:degeneracy}}
\vspace{0.3cm}

\centerline{{\vbox{\epsfxsize=8.4cm
\epsfbox{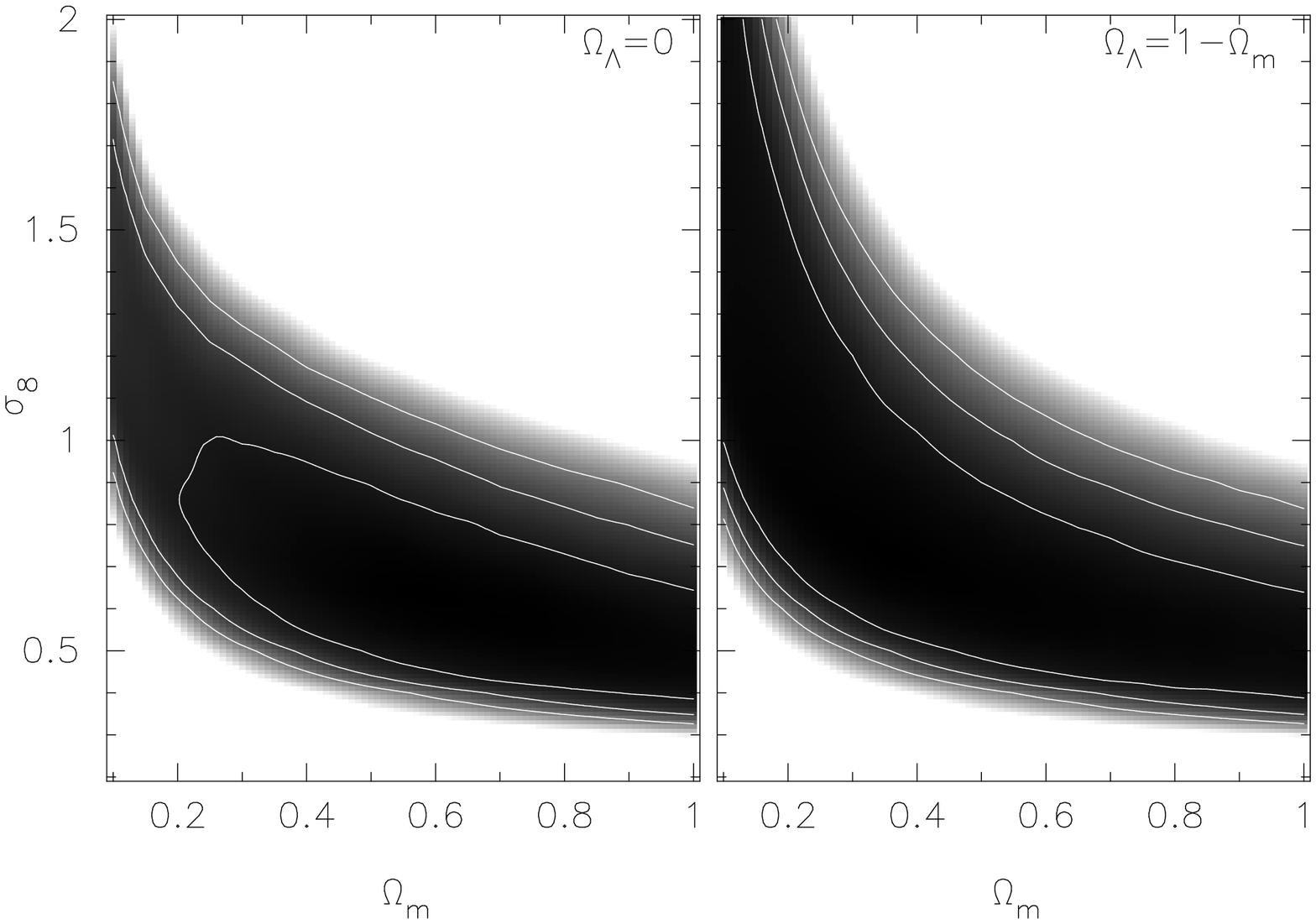}}}}
\figcaption{Constrains on $\Omega_{\rm m}$ and $\sigma_8$ 
obtained after a marginalization over $\Gamma\in[0.1,0.4]$ and 
$\bar{z}_s \in[0.6,1.4]$.
Contours indicate 68.3, 95.4 and 99.73\% confidence levels.
\label{fig:oms8_map}}
\vspace{0.3cm}

Figure \ref{fig:degeneracy} shows the degeneracy among the four
parameters.
The left panels clearly show the strong degeneracy between $\sigma_8$ and
$\bar{z}_s$ (or $\Gamma$). {}From these plots the following two points
become clear:
(i) redshift information about the source galaxies is crucial to obtaining
a tight constraint
on $\sigma_8$ (Bernardeau et al. 1997; Jain \& Seljak 1997).
(ii) A constraint on $\Gamma$ from other independent observations
(such as galaxy clustering and/or CMB anisotropies) is very useful
to break the degeneracy among the parameters.
The right panels of Figure \ref{fig:degeneracy} show the same
likelihood maps but for $\Omega_{\rm m}$. The constraints on
$\Omega_{\rm m}$ are weaker than those on $\sigma_8$, because the cosmic
shear correlation is less sensitive to
$\Omega_{\rm m}$ than $\sigma_8$ (Bernardeau et al. 1997;
Jain \& Seljak 1997).

\begin{table*}
\caption{95.4\% one-(two-)parameter confidence intervals obtained from 
the aperture mass variance,  $\Delta \chi^2 = 4.00$ $(6.17)$, of 
$\Omega_{\rm m}$ and $\sigma_8$ for different priors. (...) indicates 
that no constraint is placed within the intervals $0.1<\Omega_{\rm m}<1.0$
and $0.2<\sigma_8<2.0$.}
\label{table:confidence}
\begin{tabular}{llcccc}
\hline
\multicolumn{2}{c}{Priors} & 
\multicolumn{2}{c}{$\Omega_{\rm m}+\Omega_{\Lambda}=1$} &
\multicolumn{2}{c}{$\Omega_{\Lambda}=0$} \\
$\langle z_s \rangle$ & $\Gamma$ & $\Omega_{\rm m}$ & $\sigma_8$ &
$\Omega_{\rm m}$ & $\sigma_8$\\
\hline
$\in[0.6,1.4]$ & $\in[0.05,0.75]$ & 
... (...) & $>0.22$ (...) &  
$>0.37$ ($>0.23$) & 0.22-0.93 ($<1.03$) \\
$\in[0.6,1.4]$ & $\in[0.1,0.4]$   & 
... (...) & $>0.29$ ($>0.26$) & 
$>0.41$ ($>0.26$) & 0.29--0.79 (0.26--0.86) \\
$\in[0.6,1.4]$ & 0.21             & 
... (...) & 0.33--1.92 ($>0.31$) & 
$>0.43$ ($>0.26$) & 0.33--0.70 (0.31--0.76) \\
$\in[0.6,1.2]$ & $\in[0.05,0.75]$ & 
... (...) & $>0.24$ ($>0.21$) &  
$>0.36$ ($>0.24$) & 0.23-0.93 (0.21--1.03) \\
$\in[0.6,1.2]$ & $\in[0.1,0.4]$   & 
... (...) & $>0.31$ ($>0.29$) & 
$>0.40$ ($>0.25$) & 0.31--0.79 (0.29--0.87) \\
$\in[0.6,1.2]$ & 0.21             & 
... (...) & 0.35--1.92 ($>0.33$) & 
$>0.43$ ($>0.28$) & 0.35--0.70 (0.33--0.76) \\
$\in[0.8,1.4]$ & $\in[0.05,0.75]$ & 
... (...) & 0.21--1.65 ($<1.75$) & 
$>0.36$ ($>0.21$) & 0.21--0.76 ($<0.81$) \\
$\in[0.8,1.4]$ & $\in[0.1,0.4]$   & 
... (...) & 0.28--1.44 (0.25--1.56) & 
$>0.39$ ($>0.22$) &  0.27--0.63 (0.25--0.68) \\
$\in[0.8,1.4]$ & 0.21             & 
... (...) & 0.32--1.29 (0.31--1.39) & 
$>0.41$ ($>0.23$) & 0.32--0.55 (0.31--0.61) \\
\hline
\end{tabular}
\end{table*}

In addition to the default prior, we adopt other priors from
information obtained from other observations.
We estimate the mean redshift of our galaxy catalog using redshift
distributions from other deep surveys.
The redshift distributions of faint galaxies in the Hubble Deep Fields
North and South have been estimated using spectroscopic and photometric
redshift techniques by several groups (Fern\'{a}ndez-Soto et al.~1999;
Cohen et al.~2000; Fontana et al.~1999; 2000; Yahata et al.~2000).
Note that there is a small discrepancy between the two datasets,
which is probably due to field-to-field variation.
If we assume that our galaxy catalog has a similar redshift distribution
to the Hubble Deep Field data, the mean redshift of our catalog
($22.5<R_c<26$) is estimated to be larger than $z=1$.
On the other hand, the redshift distribution of galaxies in the Subaru
deep
survey field has been estimated using the photometric redshift technique with
$B~V~R_c~i'~z'$ data (Furusawa 2002; Furusawa et al. in preparation,
see also Kashikawa et al. 2003).
They found that the mean redshift of faint galaxies ($R>24$) is
systematically lower than the HDF data.
Note that the discrepancy is not very significant, given the large error
bars due to small-number statistics.
If we take the redshift distribution of the Subaru deep survey data,
the median redshift of our catalog can be as low as $z=0.7$.
The difference between these two estimates suggests that
the field-to-field variation can be quite large.
It is also important to take into account that the weighting of
galaxies might change the redshift distribution in a non-trivial
manner.
Considering these points, we adopt a conservative constraint of
$\bar{z}_s\in[0.6,1.4]$, and two tighter constraints of
$\bar{z}_s\in[0.6,1.2]$ and $\bar{z}_s\in[0.8,1.4]$ to determine the
impact of the source redshift information.
We constrain $\Gamma$ using the studies of galaxy
clustering by the SDSS (Dodelson et al.~2002; Szalay et al.~2001)
and the APM survey (Eisenstein \& Zaldarria 2001).
As there is a wide dispersion in $\Gamma$ values among these studies,
whereas the statistical error bars in each measurement are small,
we adopt a conservative constraint, $\Gamma\in[0.1,0.4]$,
which covers the 68\% confidence intervals of these three studies.
We also take an extreme constraint of $\Gamma=0.21$, where the
68\% confidence intervals of the three studies overlap, to
determine how well independent information on $\Gamma$ improves
the constraint on $\Omega_{\rm m}$ and $\sigma_8$.

Figure \ref{fig:oms8_map} shows the constraint on $\Omega_{\rm m}$ and
$\sigma_8$ obtained from marginalization over $\Gamma\in[0.1,0.4]$ and
$\bar{z}_s\in[0.6,1.4]$.
As expected, a strong degeneracy between $\Omega_{\rm m}$ and $\sigma_8$
is found.
The confidence intervals of $\Omega_{\rm m}$ and
$\sigma_8$ from various priors are summarized in Table
\ref{table:confidence}.
It is clear from this Table that information about the redshift
distribution of galaxies and $\Gamma$ can indeed give a tighter
constraint on $\Omega_{\rm m}$ and $\sigma_8$.
Although the current constraint is not very tight, we may
rule out the following two models.The COBE normalized high density CDM model (Bunn \& White 1997, i.e.,
$\Omega_{\rm m}=1$, $\Omega_\lambda=0$, $\sigma_8=1.2$) is ruled out
at more than the 99.9\% confidence level.
This model predicts too high an amplitude at all scales.
For the open model, low density models with $\Omega_{\rm m}<0.2$
are ruled out at more than the 68\% confidence level.
As pointed out by Van Waerbeke et al. (2002; see also Schneider et al.
1998), the incompatibility is because such low density models predict
very large power at small scales.
We obtain the following fitting function for the 95\% confidence
contours plotted in Figure \ref{fig:oms8_map},
\begin{equation}
\label{eq:fit_flat}
\sigma_8=(0.50_{-0.16}^{+0.35})\Omega_{\rm m}^{-0.37}, \quad \mbox{for }
\Omega_{\rm m}+\Omega_{\Lambda}=1,
\end{equation}
and
\begin{equation}
\label{eq:fit_open}
\sigma_8=(0.51_{-0.16}^{+0.20})\Omega_{\rm m}^{-0.34}, \quad \mbox{for }
\Omega_{\Lambda}=0.
\end{equation}

\vspace{0.3cm}
\centerline{{\vbox{\epsfxsize=8.4cm
\epsfbox{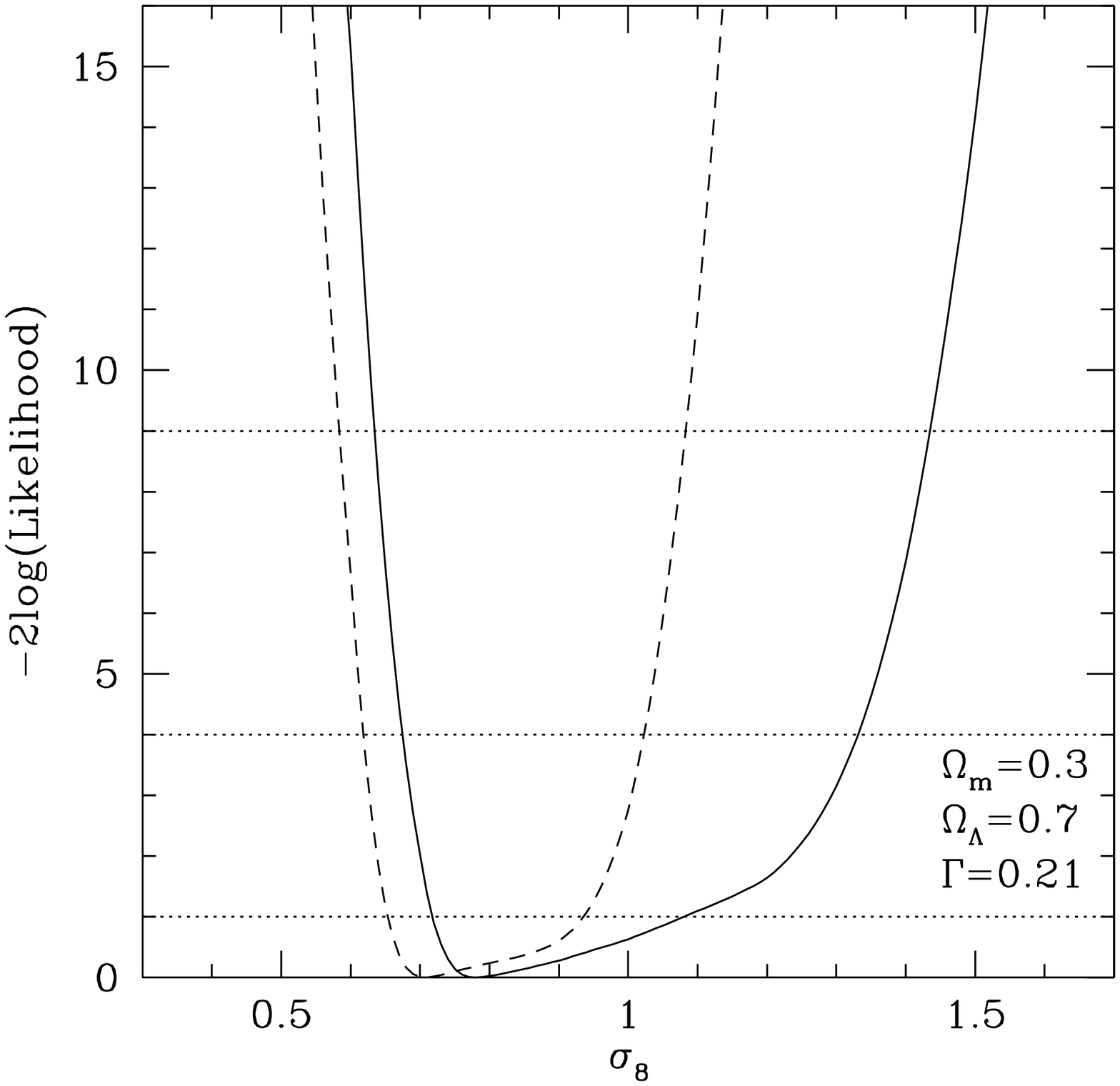}}}}
\figcaption{The one-dimensional likelihood function ($\Delta \chi^2$) 
for $\sigma_8$ in the flat $\Lambda$CDM cosmological model with 
$\Omega_{\rm m}=0.3$ ($\Omega_{\Lambda}=0.7$) and $\Gamma=0.21$
for priors $\bar{z}_s\in[0.6,1.2]$ (the solid line) and 
$\bar{z}_s\in[0.8,1.4]$ (dashed line). 
The horizontal dotted lines indicate, form lower to upper, 68.3, 95.4 and 
99.73\% confidence levels.
\label{fig:s8}}
\vspace{0.3cm}

Figure \ref{fig:s8} shows the one-dimensional likelihood function
for $\sigma_8$ in the currently popular flat $\Lambda$CDM cosmological
model with $\Omega_{\rm m}=0.3$ ($\Omega_{\Lambda}=0.7$) and
$\Gamma=0.21$.
The solid curve is for the prior $\bar{z}_s\in[0.6,1.2]$, while
the dashed curve is for $\bar{z}_s\in[0.8,1.4]$.
Specifically, the one-dimensional confidence intervals of $\sigma_8$
for the 95.4\% level are:
$0.68<\sigma_8<1.33$ for $\bar{z}_s\in[0.6,1.2]$,
$0.62<\sigma_8<1.02$ for $\bar{z}_s\in[0.8,1.4]$, and
$0.62<\sigma_8<1.32$ for $\bar{z}_s\in[0.6,1.4]$.
A strong degeneracy between $\sigma_8$ and $\bar{z}$ is evident.
Especially, the upper bound of the $\sigma_8$ confidence interval is
very sensitive to the choice of the lower limit of the mean redshift.
This is also seen in the upper left panel of Figure
\ref{fig:degeneracy}. {}From these results we may say that, for the flat
$\Lambda$CDM cosmological model, we obtain a relatively reliable
lower bound of $\sigma_8>0.62$ (95.4\% confidence)
for a reasonable choice of $\bar{z}_s$.
However, the upper limit is uncertain;
it depends strongly on the choice of the mean redshift.
A conservative conclusion is $\sigma_8<1.33$ (95.4\% confidence),
but it is tightened by $\sigma_8<1.02$ if the mean redshift is as large
as $\bar{z}_s=0.8$.

Our results are broadly consistent with constraints obtained from other
cosmic shear surveys (Maoli et al.~2001; Bacon et al.~2002;
Van Waerbeke et al.~2002; Hoekstra et al.~2002b; Refregier, Rhodes
\& Groth 2002; Brown et al.~2003; Jarvis et al.~2003).
This consistency is remarkable, given that the data have been compiled
from different instruments, filters and survey depths.
However, there is a spread of confidence intervals among the surveys.
In fact, Bacon et al.(2002) and Van Waerbeke et al. (2002) obtained
a slightly higher normalization, $\sigma_8\sim 0.95$ for the
$\Lambda$CDM model, which is incompatible with our 68.3\% confidence
interval
for $\bar{z}_s\in[0.8,1.4]$.
Note that Bacon et al.(2002) did not decompose the shear correlation
function into E and B modes. Thus, their correlation function could be
biased on the high side.
The source of the small discrepancy among the cosmic shear surveys is
unclear: it could be field-to-field variance, it could arise from the
different analysis schemes, or it could be due to a mis-choice
of the redshift distribution of galaxies.
Hirata \& Seljak (2003) investigated biases induced by the conversion
between the observed image shape to shear distortion in current weak
lensing analyses.
They found that the non-Gaussianity of the point spread function has a
significant effect and can lead to up to a 15\% error in $\sigma_8$
depending on the method of analysis.
A wider field, multi-color survey, and an analysis scheme calibrated 
using mock observations (as reported by Erben et al. 2000) are
needed to improve the accuracy of the cosmic shear analysis.

It is interesting to compare our results with the $\sigma_8$ values
obtained from the number density of rich clusters of galaxies published 
by many other groups.
It should, however, be emphasized that there is a large dispersion in
$\sigma_8$ values among these studies, whereas the errors in each
measurement are quite small.
The $\sigma_8$ values reported range from $\sigma_8 \lesssim 0.7$
(e.g., Borgani et al. 1999; 2001; Seljak 2002;
Vianna, Nichol \& Liddle 2002) to $\sigma_8 \gtrsim 0.9$ (e.g., Eke, Cole
\&
Frenk, 1996; Kitayama \& Suto 1996; 1997; Bahcall \& Fan 1998;
Pen 1998; Pierpaoli et al. 2001) for the standard $\Lambda$CDM model.

This spread may reflect mainly the uncertainty in
the relation between the mass and X-ray temperature of clusters.
If we take the prior of $\bar{z}_s\in[0.6,1.4]$, both values are well
within our 95.4\% confidence interval.
However, larger $\sigma_8$ values are not preferred by the constraint
obtained from the high-$\bar{z}_s$ prior, but are well within the
95.4\% confidence interval of the low-$\bar{z}_s$ prior.
Thus, if the mean redshift of our galaxy catalog is as high as that
estimated from the HDF data ($\bar{z}_s\gtrsim 1$), our result
is more in accord with the lower $\sigma_8$ value.

Recently, Spergel et al.~(2003) combined CMB measurements from WMAP
(Bennet et al.~2003 and references therein), CBI (Person et al.~2002) and
ACBAR (Kuo et al~2002), the galaxy power spectrum from the 2dF galaxy
redshift survey
(Percival et al.~2001; Verde et al.~2002), and the measurements of the
Lyman $\alpha$ power spectrum (Croft et al.~2002; Gnedin \& Hamilton
2002)
to find the best fit cosmological model and obtained 
$\sigma_8=0.84\pm0.04$ (68\% confidence).
This result is in good agreement with our cosmic shear constraints.

\section{Summary}

We analyzed the Suprime-Cam 2.1 deg$^2$ $R_c$-band data and measured the 
cosmic shear correlation.
Suprime-Cam has a wide field of view of $34\arcmin\times27\arcmin$,
and its superb imaging quality provides a very small RMS value of star
ellipticities of 2.8\%, which after PSF corrections is reduced to 1.0\%.
These advantages combined with the large light gathering power of the
8.2-m Subaru telescope make Suprime-Cam an almost ideal camera for a
weak lensing survey.

For the cosmic shear correlation measurement we used galaxies with
$22.5<R_c<26$ and an image size larger than the seeing size.
We detected a non-zero cosmic shear two-point correlation
function of up to 40\arcmin.
However, this result may be contaminated by shear that is not
caused by gravitational lensing.
We thus adopted the aperture mass variance, which naturally
decomposes the correlation signal into E and B modes (the latter
arises from shear whose origin is not in gravitational lensing).

We detected a non-zero E mode variance on scales from
$\theta_{ap}=2\arcmin$ to 40\arcmin.
As the aperture mass probes a scale of $\theta_{ap}/5$,
the signals come from effective scales of $0.5\arcmin < \theta
<10\arcmin$, corresponding to the quasi-linear to nonlinear regimes.
We also detected a small but non-zero B mode variance on scales
larger than $\theta_{ap}>5\arcmin$.
Currently, the origin of this B mode variance is not clear.
One possibility is an incorrect anisotropic PSF correction.
To test this possibility, we repeated the anisotropic PSF correction
using different procedures (see the Appendix for details), namely:
(i) a second order bi-polynomial fit to PSF,
(ii) a pointing-by-pointing correction without masking overlapped
regions and
(iii) using fainter stars for modeling the PSF anisotropy.
We did not find significant problems in our PSF correction procedure.

Interestingly, the amplitude of the B mode variance on larger scales is
similar to that found by Van Waerbeke et al. (2002),
though their survey depth was slightly shallower than ours.
On the other hand, Hoekstra et al.~(2002b) found a vanishing B mode
variance on scales larger than 10 arcmin in their shallow data
($R_c<24$).
These results may suggest that the current procedures for galaxy shape
correction become problematic for fainter objects.
Future detailed studies of the origin of the B mode shear are required to 
understand and suppress this possible source of residual systematic
error.
Also, a calibration of the analysis scheme using mock data is needed to
improve the accuracy of the cosmic shear analysis.

We performed a maximum likelihood analysis in a four-dimensional space
of $\sigma_8$, $\Omega_{\rm m}$, $\Gamma$ and $\bar{z}_s$.
We included three possible sources of error: the statistical noise,
the cosmic variance, and the residual systematic estimated from the B
mode variance.
We derived joint constraints on two parameters by marginalizing over
the two remaining parameters.
We obtained a weak upper limit of $\Gamma<0.5$ for $\bar{z}_s>0.9$
(68\% confidence).
We also showed that independent information on $\Gamma$ can reduce
the degeneracy among the parameters.
For the prior $\Gamma\in[0.1,0.4]$ and $\bar{z}_s\in[0.6,1.4]$,
we found
$\sigma_8=(0.50_{-0.16}^{+0.35})\Omega_{\rm m}^{-0.37}$ for
$\Omega_{\rm m}+\Omega_{\Lambda}=1$ and
$\sigma_8=(0.51_{-0.16}^{+0.29})\Omega_{\rm m}^{-0.34}$ for
$\Omega_{\Lambda}=0$ (95.4\% confidence).
Although the current constraint is not very tight, we can 
rule out the following two models:
the COBE normalized high density CDM model
($\Omega_{\rm m}=1$, $\Omega_\lambda=0$, $\sigma_8=1.2$) by
more than a 99.9\% confidence level, and
low density open models ($\Omega_{\rm m}<0.2$) by more than 68\%
confidence.
If we take the currently popular $\Lambda$CDM model ($\Omega_{\rm
m}=0.3$,
$\Omega_{\lambda}=0.7$, $\Gamma=0.21$), we obtain a one-dimensional
confidence interval on $\sigma_8$ for the 95.4\% level, 
$0.62<\sigma_8<1.32$ for $\bar{z}_s\in[0.6,1.4]$.
This result is broadly consistent with constraints from other cosmic
shear surveys and from the cluster abundance.
However, we found that the confidence interval is sensitive to the
choice of the mean redshift:
$0.68<\sigma_8<1.33$ for the prior of $\bar{z}_s\in[0.6,1.2]$,
while $0.62<\sigma_8<1.02$ for $\bar{z}_s\in[0.8,1.4]$.
The latter is incompatible with the higher $\sigma_8$ values
obtained from some cluster abundance studies.
This result clearly demonstrates that information on the redshift
distribution of the source galaxies is crucial and can significantly
tighten the confidence
interval of $\sigma_8$ and $\Omega_{\rm m}$.
The improvement of the constraint on $\sigma_8$ from the redshift
information can be estimated as follows:
the cosmic shear correlation roughly scales with $\sigma_8$ and the
mean redshift as $\xi\propto\sigma_8^{2.5}z_s^{1.5}$, thus the
uncertainly in the median redshift contributes to the error in
$\sigma_8$ as $\delta\sigma_8/\sigma_8=0.6\delta z_s/z_s$.
Therefore, the error in $\sigma_8$
due to uncertainly in the median redshift can be reduced to 10\%
by the current photometric redshift technique (e.g., Bolzonella, Miralles
\& Pell\'o 2000).

\acknowledgments
We would like to thank Y. Mellier for useful discussions and comments
on the manuscript and L. Van Waerbeke for helpful discussions about
galaxy shape analysis.
We also thank the anonymous referee for detailed and constructive
comments on an earlier manuscript, which improved the paper.
T.H. thanks K. Umetsu and T. Futamase for useful discussions.
T.H. F.N. and M.O. acknowledge support from Research Fellowships
of the Japan Society for the Promotion of Science.


\appendix
\section{Anisotropic PSF correction}
In \S 5, we found a small but non-zero B mode aperture mass variance
on scales larger than 5 arcmin.
Currently, the origin of this B mode variance is not clear.
One possibility is an incorrect anisotropic PSF correction.
To test this possibility, we repeated the anisotropic PSF correction,
but adopting different procedures:
\begin{enumerate}
\item use a second order bi-polynomial fit to
$(P_{sm}^{*})^{-1}\bm{e_{obs}^{*}}$.
\item use the pointing-by-pointing correction without masking the
overlapping regions.
\item use fainter stars for modeling the PSF anisotropy.
\end{enumerate}

First, we repeated the anisotropic PSF correction adopting a second
order bi-polynomial fit (the primary analysis uses a first order fit).
In this case, we found that both the E and B mode aperture mass variances
are almost identical to the results from our primary data, the
differences are
$1\times 10^{-6}$ at largest.
Further, higher order fits are not feasible, because of the small
number of stars in some chips.
Note that Van Waerbeke et al. (2002) reported that a higher order
polynomial fit to the PSF (third order in their case) caused a wing at
the edge of fields and produced an artificial B mode signal.

Second, we applied the anisotropic PSF correction not to each chip
separately but to each pointing, which is composed of ten chips (see
Miyazaki et al.~2002 for instrumental details of Suprime-Cam).
In this case, we did not mask the overlapping regions between different
CCD chips that result from stacking dithering exposures.
The second and fifth order bi-polynomial fits were adopted.
No significant differences were found in either the E or B mode variances between
the second and fifth order corrections.
The E mode variance is consistent with our primary data (plotted in
Figure \ref{fig:MapEB}).
The amplitude of B mode variance is also similar to the
primary data, but there is a bump at $6\arcmin<\theta_{ap}<15\arcmin$.
This scale is translated into a real scale of
$1\arcmin<\theta<2\arcmin$, which corresponds to the dithering angles
between exposures.
Thus, it is very likely that the bump arises from an inaccurate PSF
correction at the overlapping regions where the PSF anisotropy pattern
becomes very irregular\footnote{A similar B mode excess is found in
CFHT data (L. Van Waerbeke \& Y. Mellier, private communication).}.
Because of this result, we decided to mask the overlapping regions.
Also, we decided to adopt the chip-by-chip correction to avoid poor
modeling of the anisotropic PSF due to discontinuities between the
chips.

Finally, we repeated the PSF correction but adopted slightly fainter
stars to test the possibility of different responses to the PSF between
bright and faint stars.
Stars in the magnitude range $21.5<R_c<23.5$ were used for the PSF
correction with the first order fit ($20.6<R_c<23.0$ for the primary
procedure).
The number of stars selected is almost the same as the primary
selection ($\sim 1/$arcmin$^2$).
Both the E and B mode variance from these data are consistent with the
primary data (specifically the results are within the error bars of
the primary data).
A much fainter criterion for star selection gives a poor PSF model
because of contamination by small galaxies. Thus, it gives a very poor
PSF correction.

In conclusion, as far as can be determined from the tests, we did not
find a significant problem with our PSF correction procedure.
A future detailed examination of the PSF correction method
should be carried out using realistic simulation data, similar to
the exercise performed by Erben et al. (2001).
These tests, however, are beyond the scope of this paper and will be
reported elsewhere.

\end{document}